\documentclass[sigconf]{acmart}
\AtBeginDocument{%
  }

\copyrightyear{2026}
\acmYear{2026}
\setcopyright{cc}
\setcctype{by}
\acmConference[KDD '26]{Proceedings of the 32nd ACM SIGKDD Conference on Knowledge Discovery and Data Mining V.2}{August 09--13, 2026}{Jeju Island, Republic of Korea}
\acmBooktitle{Proceedings of the 32nd ACM SIGKDD Conference on Knowledge Discovery and Data Mining V.2 (KDD '26), August 09--13, 2026, Jeju Island, Republic of Korea}
\acmDOI{10.1145/3770855.3817489}
\acmISBN{979-8-4007-2259-2/2026/08}

\usepackage{makecell}
\usepackage{multirow}
\usepackage{pifont}
\usepackage[most,skins,theorems]{tcolorbox}
\usepackage{booktabs}
\usepackage{placeins}
\usepackage{tabularx}
\usepackage{array}
\usepackage[normalem]{ulem}

\newcommand{\bench}{\textsc{SciHorizon-Gene}}

\begin{document}

\title{SciHorizon-Gene: Benchmarking LLM for Life Sciences Inference from Gene Knowledge to Functional Understanding}
\renewcommand{\shorttitle}{SciHorizon-Gene}


\author{Xiaohan Huang}
\affiliation{%
  \institution{Computer Network Information Center, Chinese Academy of Sciences}
  \institution{University of the Chinese Academy of Sciences}
  % \institution{CNIC, CAS}
  % \institution{UCAS}
  \city{Beijing}
  \country{China}}
\email{xhhuang@cnic.cn}

\author{Meng Xiao}
\affiliation{
  \institution{Computer Network Information Center, Chinese Academy of Sciences}
  \city{Beijing}
  \country{China} \\
  \institution{DUKE-NUS Medical School, National University of Singapore}
  \city{Singapore}
  \country{Singapore}}
\authornote{Corresponding author: Meng Xiao (shaow@cnic.cn)}
\email{shaow@cnic.cn}
% \author{Chuan Qin \\ Qingqing Long}
% \affiliation{%
%   \institution{Computer Network Information Center, CAS}
%   \institution{University of Chinese Academy of Sciences}
%   \city{Beijing}
% \email{chuanqin0426@gmail.com}

\author{Chuan Qin}
\affiliation{%
  \institution{Computer Network Information Center, Chinese Academy of Sciences}
  \institution{University of the Chinese Academy of Sciences}
  \city{Beijing}
  \country{China}
}
\email{chuanqin0426@gmail.com}

\author{Qingqing Long}
\affiliation{%
  \institution{Computer Network Information Center, Chinese Academy of Sciences}
  \institution{University of the Chinese Academy of Sciences}
  \city{Beijing}
  \country{China}
}
\email{qqlong@cnic.cn}

\author{Jinmiao Chen}
\affiliation{%
  \institution{DUKE-NUS Medical School, National University of Singapore}
  \institution{Singapore Immunology Network, Agency for Science, Technology and Research}
  \city{Singapore}
  \country{Singapore}}
\email{micchenj@nus.edu.sg}

\author{Yuanchun Zhou}
\affiliation{%
  \institution{Computer Network Information Center, Chinese Academy of Sciences}
  \institution{University of the Chinese Academy of Sciences}
  \city{Beijing}
  \country{China}}
\email{zyc@cnic.cn}

\author{Hengshu Zhu}
\affiliation{%
  \institution{Computer Network Information Center, Chinese Academy of Sciences}
  \institution{University of the Chinese Academy of Sciences}
  \city{Beijing}
  \country{China}}
\email{hszhu@cnic.cn}

\renewcommand{\shortauthors}{Xiaohan Huang et al.}

% \begin{abstract}
%   A clear and well-documented \LaTeX\ document is presented as an
%   article formatted for publication by ACM in a conference proceedings
%   or journal publication. Based on the ``acmart'' document class, this
%   article presents and explains many of the common variations, as well
%   as many of the formatting elements an author may use in the
%   preparation of the documentation of their work.
% \end{abstract}
\begin{abstract}
Large language models (LLMs) have shown growing promise in biomedical research, particularly for knowledge-driven interpretation tasks. 
However, their ability to reliably reason from gene-level knowledge to functional understanding, a core requirement for knowledge-enhanced cell atlas interpretation, remains largely underexplored.
To address this gap, we introduce \bench, a large-scale gene-centric benchmark constructed from authoritative biological databases.
The benchmark integrates curated knowledge for over 190K human genes and comprises more than 540K questions covering diverse gene-to-function reasoning scenarios relevant to cell type annotation, functional interpretation, and mechanism-oriented analysis.
Motivated by behavioral patterns observed in preliminary examinations, \bench\ evaluates LLMs along four biologically critical perspectives: research attention sensitivity, hallucination tendency, knowledge completeness, and literature influence, explicitly targeting failure modes that limit the safe adoption of LLMs in biological interpretation pipelines.
We systematically evaluate a wide range of state-of-the-art general-purpose and biomedical LLMs, revealing substantial heterogeneity in gene-level reasoning capabilities and persistent challenges in generating faithful, complete, and literature-grounded functional interpretations. 
Our benchmark establishes a systematic foundation for analyzing LLM behavior for gene knowledge inference and functional understanding, which offers insights for model selection and development, with direct relevance to knowledge-enhanced biological interpretation. 
\sloppy
Our dataset is publicly available at \url{https://www.scidb.cn/detail?dataSetId=4700d275bd5741958894d3739cbdc1dd\&version=V1}.
\end{abstract}



% \keywords{Do, Not, Us, This, Code, Put, the, Correct, Terms, for,


\maketitle

\begin{figure*}[!ht]
    \centering
    \includegraphics[width=0.95\linewidth]{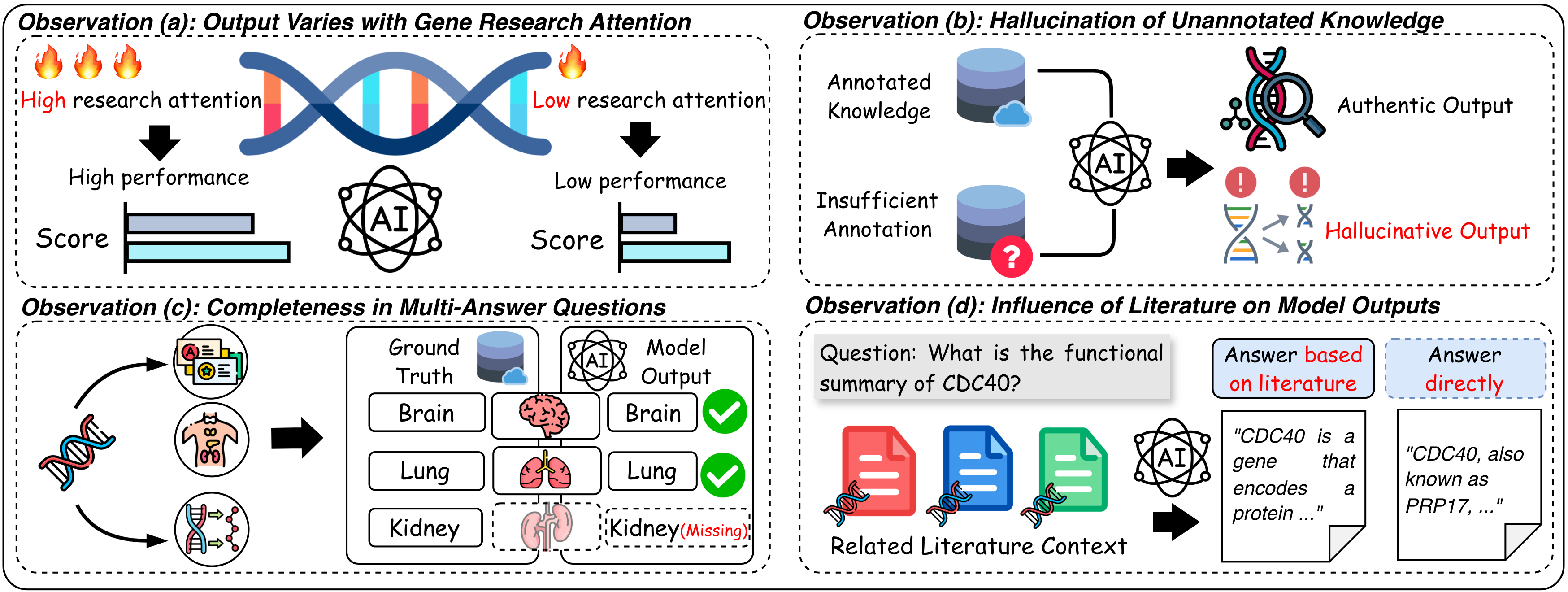}
    \caption{Observations of LLM behavior on gene-related tasks, motivating the need for our gene-centric benchmark. (a) Model performance differs across genes with varying levels of research attention. (b) Models tend to yield hallucinated answers when responding to genes with insufficient annotation. (c) LLMs often produce partially correct but incomplete answers in multi-answer questions. (d) Model outputs vary notably depending on whether the relevant literature context is provided.}
    % \vspace{-0.3cm}
    \label{fig:motivation}
\end{figure*}

\section{Introduction}
Recent advances in large language models(LLMs) have substantially expanded their role in scientific research~\cite{zhang2025exploring,wang2023scientific,reddy2025towards,luo2025large}, giving rise to the broader paradigm of AI for Science~\cite{qin2025scihorizon,dagdelen2024structured,DiffCR2024Zou}.
Within this landscape, the biomedical domain has emerged as a prominent application area, where LLMs are increasingly applied to knowledge-intensive tasks such as medical question answering~\cite{singhal2023large,ye2023needed,singhal2025toward,wang2025survey,xiao2025m} and large-scale literature understanding~\cite{zhou2025large,luo2022biogpt,chen2024genesum,cai2023resolving,cai2025knowledge,wang2024biorag}. 
Beyond classical QA and summarization, core biomedical workflows increasingly demand automated interpretation and integration of complex experimental data, exemplified by reference mapping and functional annotation in next-generation cell atlas studies~\cite{li2024screader,liu2025gut}. 
These tasks are not merely extensions of natural language processing~\cite{xiao2023hierarchical,xiao2021expert,chen-etal-2026-gendis, xu-etal-2026-tlsa, qin-etal-2026-bolt}, but require reliable, function-oriented reasoning grounded in biological knowledge~\cite{xiao2025knowledge,cui2024automated}, in order to scale analysis and reduce reliance on labor-intensive manual curation. 
Among diverse forms of biological information, gene-level knowledge serves as the most fundamental abstraction, as genes constitute the basic units underlying cellular functions, disease mechanisms~\cite{jackson2018genetic}, and molecular pathways~\cite{lappalainen2024genetic}. 
Consequently, \textbf{a robust ability to reason from gene-centric knowledge to functional understanding} is essential for enabling LLMs to support meaningful, reliable, and interpretable biomedical analysis. 

In response to growing interest in biomedical applications of LLMs, a range of methods and benchmarks have been proposed to evaluate model performance in domain-specific settings. 
Existing approaches can be broadly categorized into literature-centered question answering and reasoning benchmarks, which assess models' ability to retrieve, synthesize, and reason over biomedical texts~\cite{chen2025benchmarking}, such as PubMedQA~\cite{jin2019pubmedqa}, MedMCQA~\cite{pal2022medmcqa}, and BioASQ \cite{krithara2023bioasq}, as well as instruction- or task-oriented evaluation frameworks that probe general biomedical understanding through curated prompts or examination-style questions~\cite{jin2021disease, pal2022medmcqa}. 
In parallel, recent efforts have begun to explore gene-related evaluation settings, aiming to assess LLM behavior on genomics-oriented queries and factual recall~\cite{shang2025benchmarking}.
While these methods provide valuable insights into biomedical language understanding, they largely operate at the level of documents, questions, or task modules, and do not explicitly examine how models translate gene-level knowledge into coherent functional interpretations under varying biological contexts.

Despite these advances, gene-level information poses unique challenges for LLM-based understanding. 
Gene annotations span heterogeneous attributes, including nomenclature, genomic location, molecular function, expression patterns, and disease associations, and the available evidence is highly uneven across genes. 
Well-studied genes are supported by rich curated records and dense literature, whereas many others remain sparsely annotated or appear only as provisional identifiers. 
Moreover, genomic knowledge is distributed across both structured databases and unstructured publications, making it non-trivial for LLMs to integrate evidence and maintain consistent behavior across the gene space. 
Existing evaluations of biomedical LLMs primarily focus on document-level question answering, summarization, or exam-style clinical problems, and do not systematically characterize how models understand genes with diverse attributes and varying levels of characterization. 
As a result, the gene-centric capabilities of LLMs remain insufficiently understood, both in terms of \textbf{overall performance} and \textbf{behavioral variability} across different classes of genes.

To better understand how LLMs behave on gene-related questions in practice, we conducted preliminary observations on model outputs. 
As illustrated in Figure~\ref{fig:motivation}, these observations reveal several recurring behavioral patterns: 
(a) model performance varies markedly across genes with different levels of research attention; 
(b) hallucinated descriptions frequently arise for genes with sparse or incomplete annotation; 
(c) responses to multi-answer questions are often incomplete; and 
(d) the presence or absence of literature context can substantially alter model outputs for the same query. 
We refer to these systematic tendencies as \emph{model-level behaviors}. 
Rather than focusing on isolated question-level accuracy, we investigate how models behave across genes under specific conditions. 
These observations motivate the need for evaluation settings that can systematically capture gene-centric behavioral patterns.

Building on these insights, we present a large-scale gene-centric benchmark, \bench, grounded in trusted biological data sources. 
We integrate curated information for over 190K human genes from authoritative databases and automatically generate more than 540K gene-related questions spanning multiple question types. 
The benchmark covers three core scenarios, including gene nomenclature, biological knowledge, and the influence of reference context. \bench\ incorporates four evaluation perspectives that directly reflect the challenges identified above: research attention, hallucination tendency, knowledge completeness, and literature influence. 
All metrics are automatically computable, enabling scalable and fully reproducible evaluation without manual scoring. 
We evaluate 27 open-source and proprietary models, encompassing both general-purpose and domain-specialized systems, and provide a comprehensive characterization of LLM behavior under gene-centric settings.
The results can be accessed on the SciHorizon platform under the Vertical category\footnote{https://scihorizon.cn/verticalCategory/SciHorizonGene}.
The code and evaluation prompts are provided on GitHub~\footnote{https://github.com/CNIC-DSL/SciHorizonGene}.
See Appendix~\ref{appendix:online} for more details about the online platform and open-source resources.

\section{Related Work}

In recent years, LLMs have been widely applied in various domains, including natural language processing~\cite{dong2024temporal,chen2025beyond, tong2025missteps, shen2026prompting, qin2025cotr, jiang2024enhancing,wang2026face}, recommendation systems~\cite{zhou2025enhancing,lin2024gume}, and management science~\cite{qin2025comprehensive}. 
These advances have further stimulated growing interest in their applications to biomedical research.
In this section, we first review representative biomedical LLMs developed to enhance domain-specific understanding~\cite{zhang2025comprehend} and scientific reasoning~\cite{xiao2025reinforcement}.
We then discuss existing biomedical benchmarks for evaluating LLMs.

\subsection{LLMs for Biomedical Studies}
Recent LLMs have demonstrated strong capabilities in scientific tasks~\cite{zhang2025exploring, wang2023scientific, reddy2025towards}, supported by their broad knowledge coverage and advanced language understanding abilities.
Within the biomedical domain, a growing number of specialized LLMs~\cite{singhal2023large,singhal2025toward} have been developed to better capture domain-specific terminology and conceptual structures.
BioGPT~\cite{luo2022biogpt} represents an early effort to pretrain a generative Transformer on large-scale biomedical literature, enabling improved scientific text generation.
PMC-LLaMA~\cite{wu2024pmc} incorporates medical instruction tuning on a knowledge-aware dataset to enhance its ability to follow clinically relevant prompts.
BioMistral~\cite{labrak:hal-04621178} extends this direction through pretraining on PubMed Central corpora, improving literature-grounded understanding and inference.
MedGemma~\cite{sellergren2025medgemma} exhibits strong medical comprehension, achieving performance that surpasses models of comparable size and approaches that of task-specific systems.
Biomedical-LLaMA~\cite{Bio-Medical-Llama-3-8B} and MedAlpaca~\cite{han2023medalpaca} similarly focus on biomedical applications, using domain adaptation or instruction tuning to improve performance on clinical and scientific tasks.
Together, these developments reflect the increasing interest in adapting LLMs to specialized biomedical settings.
However, their capacity to understand gene-specific knowledge at a fine granularity remains insufficiently explored.

% \vspace{-0.2in}
\subsection{Biomedical Benchmarks for LLMs}
Existing benchmarks for biomedical LLMs primarily evaluate models on tasks involving literature-based question answering, document retrieval, sentence classification, and biomedical concept linking. 
PubMedQA~\cite{jin2019pubmedqa} constructs a biomedical question answering dataset with expert annotations, requiring models to reason over PubMed abstracts to answer research-level questions. 
MedQA~\cite{jin2021disease} collects multilingual professional medical board examination questions to assess diagnostic reasoning in open-domain clinical scenarios. 
MedMCQA~\cite{pal2022medmcqa} expands this line of work with large-scale multiple-choice questions spanning over 2.4k health topics and 21 medical subjects. 
While these benchmarks have played an important role in advancing biomedical language understanding, they are largely designed around general biomedical text processing and clinical reasoning, and do not explicitly target the fine-grained biological knowledge required to evaluate LLMs at the level of individual genes. 
In contrast to downstream biological analysis tasks, such as functional interpretation and reference-based annotation in large-scale cell atlas studies, these benchmarks typically operate at the level of documents, questions, or high-level medical concepts, without probing gene-centric reasoning behaviors.
GeneTuring~\cite{shang2025benchmarking} represents an initial effort to benchmark LLMs on genomics-related question answering by defining 16 genomics tasks with 1,600 curated questions and manually scoring 48,000 model answers.  
It manually evaluates multiple aspects of model responses, including accuracy, hallucination-related behavior, and incapacity awareness. 
However, its reliance on manual assessment limits scalability and reproducibility, making it challenging to extend evaluation to larger gene sets, more diverse contexts, or broader model families. 
Moreover, GeneTuring reports performance at the task or module level primarily, rather than explicitly structuring evaluation around systematic behavioral perspectives under varying gene- and context-specific conditions.
In contrast, our benchmark introduces an automated and multidimensional evaluation framework that supports scalable and reproducible assessment, enabling systematic characterization of LLM behavior in gene-centered settings relevant to knowledge-driven biological interpretation.

\begin{figure*}[!ht]
\centering
\includegraphics[width=0.97\linewidth]{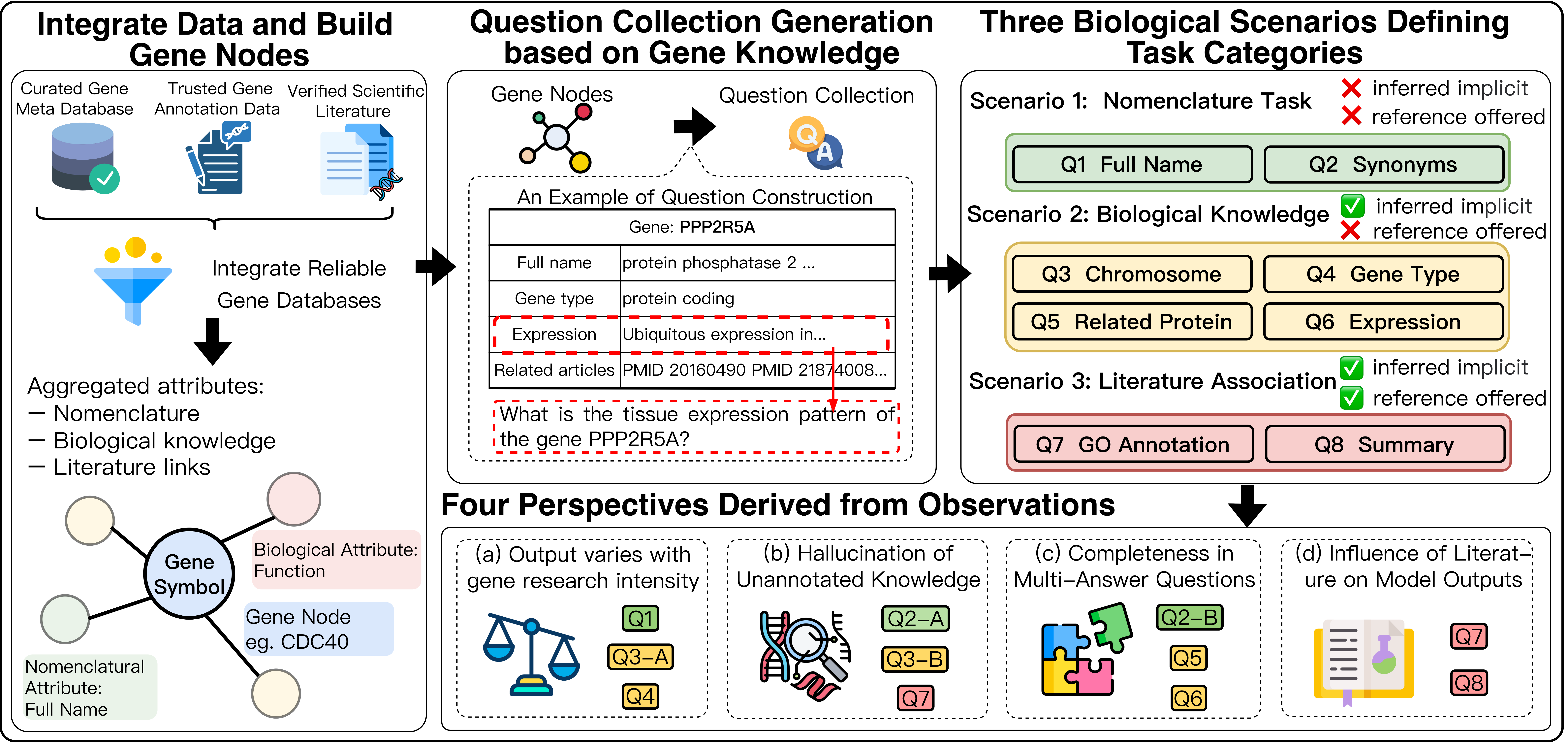}
\vspace{-0.3cm}
\caption{The benchmark integrates curated biological databases and verified literature sources to construct gene nodes.
These nodes provide a unified representation of gene attributes and support the generation of diverse, knowledge-grounded question sets.
The question collections are categorized into three biological task scenarios.
LLM performance is subsequently evaluated from four perspectives that reflect key behavioral patterns.}
\vspace{-0.4cm}
\label{fig:method}
\end{figure*}
\section{Benchmark Construction Pipeline}
This section describes the overall pipeline of our benchmark.
As shown in Fig~\ref{fig:method}, we begin by collecting and integrating publicly available datasets to construct a gene-centric knowledge resource.
Based on the integrated data, we then generate questions spanning three biological scenarios.
Finally, we introduce four curated evaluation perspectives for systematically assessing LLM behaviors.
% \vspace{-0.2in}
\subsection{Data Sources and Gene-Centric Integration}
We collected curated metadata for human genes, including standardized nomenclature and other genomic information from NCBI Gene~\cite{ncbi_gene}. 
From the Gene Ontology (GO) database~\cite{gene_ontology}, we obtained structured functional annotations that describe biological processes, molecular functions, and cellular components. 
To incorporate literature-based context, we retrieved PubMed~\cite{pubmed} records linked to human genes, along with their corresponding abstracts. 
These complementary sources collectively provide curated gene descriptions, functional annotations, and literature evidence that support the construction of a unified gene-centric knowledge base.
We then integrate data collected from multiple sources by linking all information to their corresponding gene entities.
Let $G = \{ g_1, g_2, \ldots, g_n \}$ denote the set of human gene entities collected from the metadata. 
Each gene \( g_i \in G \) is associated with a set of structured attributes
$A_i = \{ a_{i1}, a_{i2}, \ldots, a_{i m} \},$
where each attribute corresponds to a specific category of gene information. 
For each gene \( g_i \), we additionally construct a literature reference set 
$R_i = \{ r_{i1}, r_{i2}, \ldots, r_{i k} \}$,
where each element denotes a PubMed identifier associated with the gene. 
This unified representation yields a structured and gene-centric knowledge base that forms the foundation for question generation and downstream evaluation.

\subsection{Question Collection Generation}

Based on the integrated gene-centric data, we generate benchmark questions from the attributes associated with each gene. 
For a gene \( g_i \) with attribute set \( A_i \), we formulate questions that reflect the underlying biological information. 
The question format is determined by the cardinality of the attribute value.  
Let \( |a_{ij}| \) denote the number of valid values associated with attribute \( a_{ij} \).

\begin{itemize}
    \item If \( |a_{ij}| = 1 \), we construct a single-answer question.
    \item If \( |a_{ij}| > 1 \) and \( |a_{ij}| \) is finite, we generate a multiple-answer question.
    \item If \( |a_{ij}| = \infty \), corresponding to open-ended textual attributes such as functional descriptions, we formulate a generative question requiring free-text responses.
    \item If \( |a_{ij}| = 0 \), indicating no recorded value in the data, the expected output is explicitly designated as "no right answer".
\end{itemize}
% For attributes linked to scientific literature, we optionally incorporate the PubMed abstracts in \( R_i \) into the open-end question context to evaluate whether models can effectively leverage abstract information.  
For attributes linked to scientific literature, we incorporate PubMed abstracts in \( R_i \) into the open-ended question context, in order to assess whether the inclusion of abstract influences the outputs.
% Examples of each question type are provided in Appendix~\ref{appendix:example&prompts}. 
This generation process results in a diverse set of questions that systematically evaluate different aspects of gene-related knowledge.

\subsection{Three Biological Scenarios}
We categorize generated questions into three biological scenarios based on how LLMs are expected to derive their answers.
The first scenario contains tasks about gene full names and synonyms, whose answers may be inferred from surface-level lexical cues. 
The second scenario includes questions on genomic functional knowledge.
These tasks cannot be resolved from naming cues alone and instead require the model to rely on internalized biological knowledge. 
The third scenario comprises questions that connect genes to gene ontology annotations and functional summaries, for which PubMed abstracts associated with the gene may be provided to supply relevant context. 
These scenarios define the task settings used for analyzing model behavior in the following sections.

\subsection{Four Evaluation Perspectives}
\begin{figure}[ht]
\centering
\includegraphics[width=0.8\linewidth]{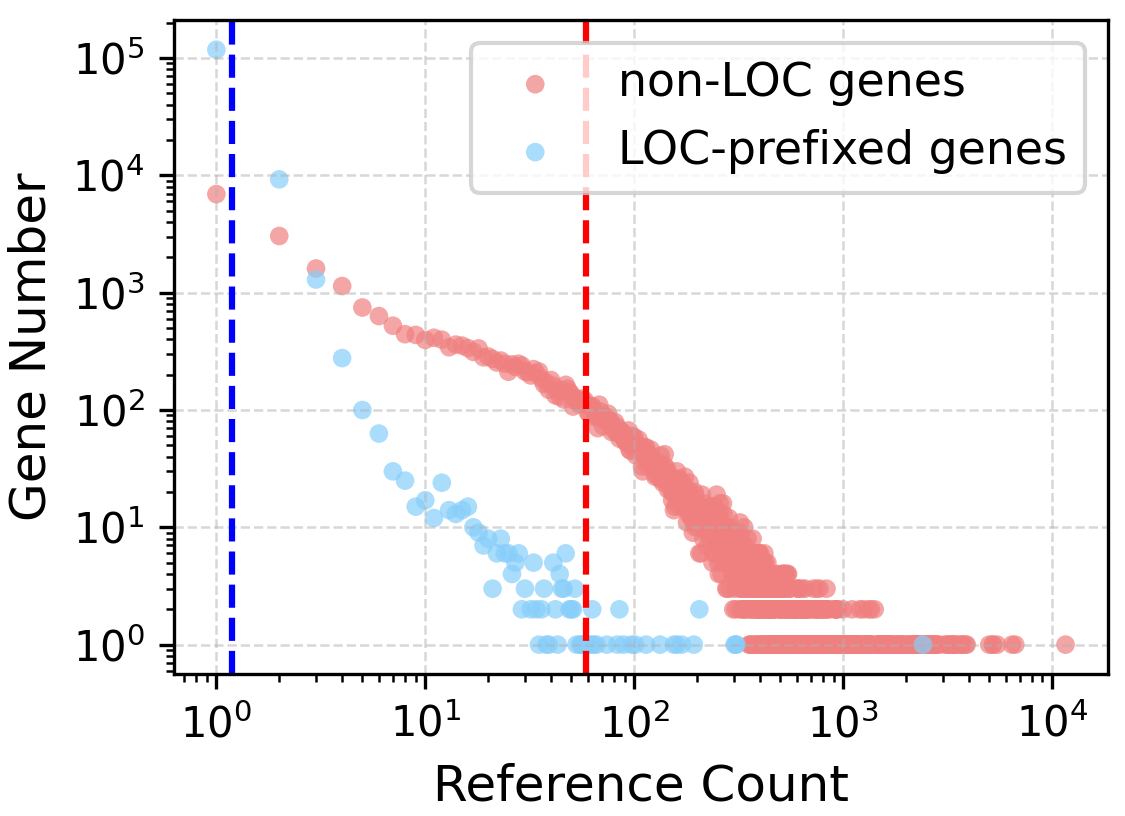}
    \caption{PubMed reference count distribution for human genes. Dashed lines mark the mean reference counts.}
    \label{fig:ref_count_distribution}
\end{figure}
We introduce four evaluation perspectives that analyze how LLMs handle challenges encountered in genomic questions.

\textbf{Research Attention.}
Genes vary substantially in the amount of available scientific literature, which serves as an indicator of research attention.
Gene names beginning with "LOC" denote provisionally annotated loci representing predicted genes that have not yet been assigned standard gene symbols.
As shown in Figure~\ref{fig:ref_count_distribution}, gene symbols begin with the prefix "LOC" typically have few associated PubMed references, whereas non-LOC genes show higher reference counts. 
The LOC group therefore provides a representation of low research attention. 
We compare model outputs between LOC-prefixed and non-LOC genes to assess whether model performance is sensitive to differences in research attention. 
This perspective enables us to characterize systematic performance differences that arise under varying levels of available biological knowledge.

\textbf{Hallucination Tendency.}
In our integrated gene-centric dataset, some attributes contain no recorded values because the corresponding biological information is not available in curated sources. 
For these cases, the correct output is explicitly defined as "no right answer".
This setting allows us to assess whether a model can recognize the absence of valid evidence and avoid generating unsupported content. 
Producing a statement when no authoritative annotation exists is considered a hallucination. 
This perspective therefore evaluates the model's reliability in scenarios where biological knowledge is currently unknown.

\textbf{Knowledge Completeness.}
Knowledge completeness characterizes an LLM's ability to cover the relevant knowledge of a given gene.
Some gene attributes, such as related proteins or expression patterns, are associated with multiple valid values.
We construct multi-answer questions from these attributes, where each question is paired with a curated answer set.
To assess model performance on knowledge completeness, we compare the model output with the gold answer set.
We quantify the proportion of gold items contained in the response.
Higher coverage indicates stronger knowledge completeness and a more thorough grasp of gene-level knowledge.

\textbf{Literature Influence.}
For Gene Ontology annotations and functional summaries, scientific publications provide essential support or context. 
In these cases, we optionally include the PubMed abstracts associated with the gene in the question context.
To assess the literature influence on model outputs, we present each question in two forms: one with the relevant abstracts provided and one without external context. 
By comparing the two responses, we evaluate whether the model can make effective use of the supplied literature and whether its answers change appropriately when additional evidence is available. 
This perspective characterizes the model's behavior under external scientific context and its ability to incorporate relevant information.

\section{Experimental Settings}
% 实验题量
% 实验框架（闭源api，开源遵守作者给的说明；pydantic格式控制）
% 选择题：cycle-permutation避免选项影响

In this section, we present the benchmark details, evaluation protocol, and metrics used in our experiments.
\vspace{-0.1 in}
% are provided in the Appendix.
\subsection{Benchmark and Evaluation Details}
% The benchmark is designed to capture diverse gene-centric behaviors under four evaluation perspectives, resulting in a large and heterogeneous set of biological questions. 
The benchmark comprises over 540K questions derived from more than 190K human genes, covering multiple task formats including single-choice, multiple-choice, and open-form answering.
We evaluate 27 LLMs, including both general-purpose and biomedical domain-specialized models. 
General-purpose models are selected based on their widespread adoption and demonstrated performance in scientific reasoning tasks, while domain-specific models are included to assess the impact of biomedical pretraining on gene-centered understanding.
% Detailed implementation procedures are provided in Appendix~\ref{appendix:evaluation-protocol}.
% Detailed benchmark statistics, model configurations, and example prompts are provided in Appendix~\ref{appendix:benchmark-details} (Table~\ref{tab:benchmark}, Table~\ref{tab:bench_details_from_knowledge} and Table~\ref{tab:models}) and Appendix~\ref{appendix:example&prompts}.
% \subsection{Evaluation Protocol}
% To mitigate positional bias in choice questions~\cite{wang2024large}, we adopt a cyclic permutation strategy~\cite{zheng2023large} that rotates the correct option across all possible positions, ensuring that evaluation reflects semantic understanding rather than positional preference.
% We sample a representative subset of 2{,}710 questions spanning all biological scenarios and evaluation perspectives, resulting in 9{,}010 evaluation instances per model.
% All evaluations are conducted in a zero-shot setting, where each query is issued independently to avoid contextual or cross-sample memory effects.
% Evaluation follows a standardized and automated pipeline designed to ensure scientific rigor and reproducibility.
% Implementation details are described in Appendix~\ref{appendix:evaluation-protocol}.
To mitigate positional bias in choice questions~\cite{wang2024large}, we use a cyclic permutation strategy~\cite{zheng2023large} that rotates the correct option across positions. 
We sample 2{,}710 representative questions covering all biological scenarios and evaluation perspectives, yielding 9{,}010 evaluation instances per model. 
All evaluations are conducted in a zero-shot setting, with each query issued independently to avoid contextual or cross-sample memory effects. 
All models are evaluated using their default or officially recommended settings.
Detailed benchmark statistics are provided in Appendix~\ref{appendix:benchmark-details}. 
The evaluation follows a standardized automated pipeline for rigor and reproducibility, with implementation details provided in Appendix~\ref{appendix:evaluation-protocol}.
% Our evaluation data can be accessed through Science Data Bank~\footnote{https://www.scidb.cn/detail?dataSetId=4700d275bd5741958894d3739cbdc1dd\&version=V1}, with all related online resources detailed in the Appendix.
% Detailed descriptions of all evaluation-related resources are provided in the Appendix~\ref{appendix:online}.

\subsection{Evaluation Metrics}

We adopt automatically computable metrics tailored to each evaluation perspective and question type.
For research attention and hallucination tendency, accuracy on single-choice questions is used, where hallucination is explicitly defined as selecting any substantive option when no correct answer exists.
Knowledge completeness is assessed on tasks with multiple valid answers using macro-averaged precision, recall, and F1.
For functional summary tasks, we measure semantic similarity using ROUGE-L~\cite{lin2004rouge} and BERTScore~\cite{zhang2019bertscore}, and additionally track perplexity and answer length to characterize changes in fluency and verbosity.
To assess literature influence in Gene Ontology prediction, we adapt the CAFA evaluation protocol~\cite{jiang2016expanded} to LLM outputs.
Formal definitions and computation details are provided in Appendix~\ref{appendix:metrics}.

\section{Results}
% \input{table/main_table}
% 总结
% In this section, we aim to address the following research questions regarding LLMs in gene knowledge evaluation:\\
% (1) How well do LLMs comply with task-specific format instructions?\\
% (2) Can LLMs capture and reason over human gene knowledge?\\
% (3) Do LLMs perform differently in reasoning about LOC-annotated versus well-annotated genes?\\
% (4) Can LLMs suppress hallucinations in gene-related question?\\
% (5) How comprehensively do LLMs cover multiple valid answers within gene-related knowledge?\\
% (6) Does reference abstract always improve the performance?
%  Do domain-specific models demonstrate deeper understanding of gene knowledge?\\
% (7) Does model evolution and scaling improve gene knowledge understanding?\\
% (8) What characteristics distinguish the top-performing models?\\
In this section, we first analyze LLM behavior along four evaluation perspectives: research attention (Sec~\ref{result:research_attention}), hallucination tendency (Sec~\ref{result:hallucination}), knowledge completeness (Sec~\ref{result:completeness}), and literature influence (Sec~\ref{result:influence_literature}), followed by a focused comparison of the top-performing models (Sec~\ref{result:top_models}).
We then summarize model performance across different genomic knowledge (Sec~\ref{result:overall}). 
% Detailed numerical results under four evaluation perspectives are provided in Appendix~\ref{appendix:detailed_results}.
% Detailed results under four evaluation perspectives are available on the SciHorizon platform\footnote{https://scihorizon.cn/verticalCategory/SciHorizonGene}.

\subsection{Research Attention and Uneven Gene Understanding}
\label{result:research_attention}
\begin{figure}[!ht]
\centering
\includegraphics[width=0.9\linewidth]{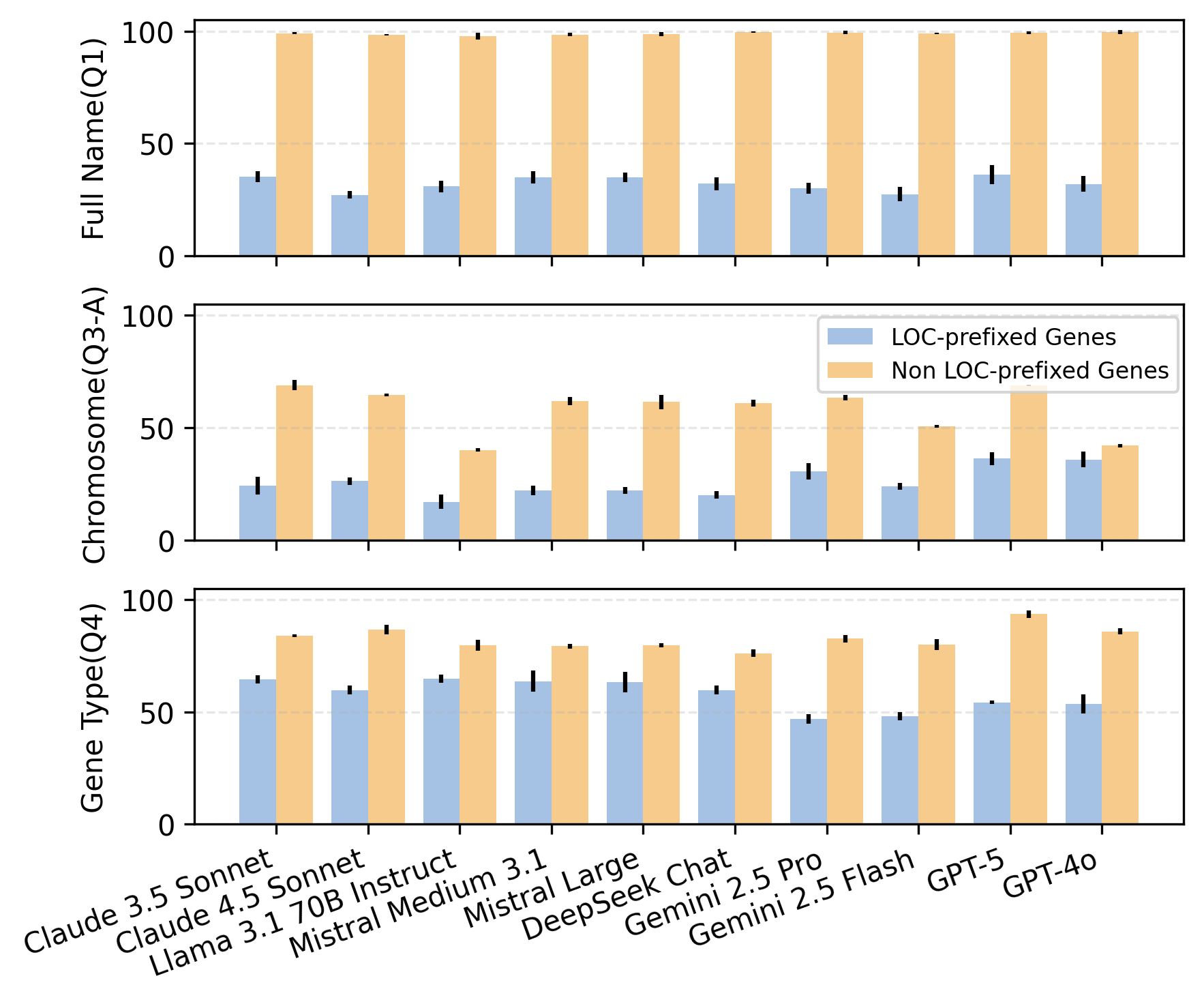}
\vspace{-0.2cm}
    \caption{Model performance on three tasks for high- and low-research attention genes. All tasks are single-choice, and accuracy is used as the evaluation metric.}
    \vspace{-0.4cm}
    \label{fig:study_bias}
\end{figure}
In this experiment, we answer the question: \textit{Do LLMs perform differently when handling genes with high versus low research attention?}
Figure~\ref{fig:study_bias} reports model accuracy across three tasks and shows a consistent pattern: models achieve substantially higher accuracy on questions associated with high-attention genes than on those involving low-attention genes.
A plausible explanation for this pattern is that high-attention genes are more frequently mentioned in biomedical literature, increasing their exposure during model pretraining and leading to stronger performance.  
In addition, these genes often carry semantically meaningful names that provide lexical cues useful for inference, whereas low-attention genes are primarily represented by alphanumeric identifiers that offer little linguistic signal.  
These findings highlight a pronounced research attention sensitivity, reflecting uneven understanding of gene knowledge across the research attention.

% hallucination
\subsection{Hallucination Resistance on Questions without Valid Answers}
\label{result:hallucination}
\begin{figure}[ht]
\centering
\includegraphics[width=0.8\linewidth]{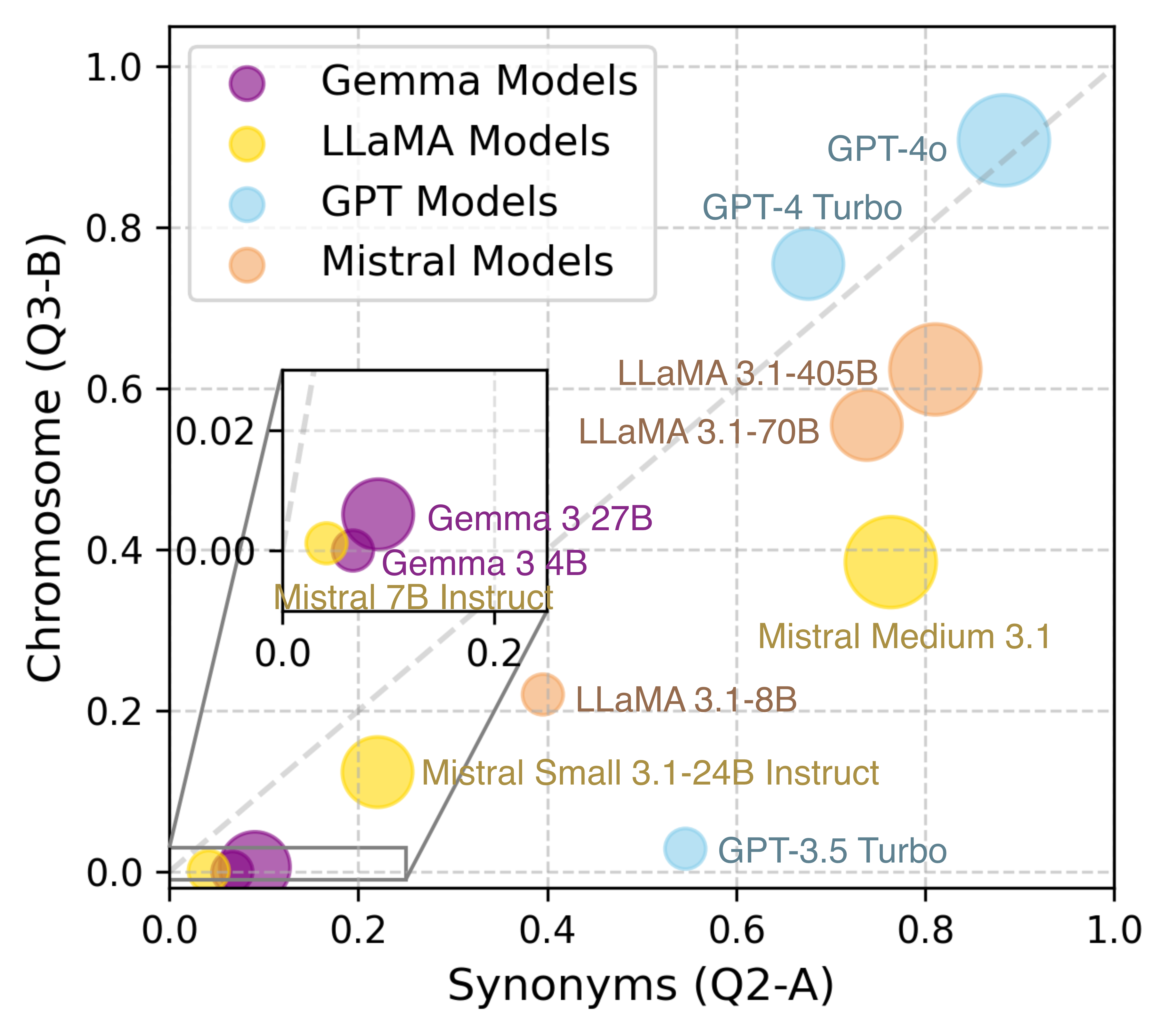}
    \caption{Hallucination resistance of LLMs across Synonyms and Chromosome questions. All tasks are single-choice, and accuracy is used as the evaluation metric. Larger markers qualitatively indicate larger or newer LLMs.}
    \label{fig:hallucination}
    \vspace{-0.4cm}
\end{figure}
This part seeks to answer the question: \textit{Can LLMs suppress hallucinations when gene-related questions have no valid answer?}
We evaluate models on synonym and chromosome location questions whose correct option is defined as "no right answer".
Figure~\ref{fig:hallucination} plots hallucination resistance for models with accuracy on synonym and chromosome questions. 
A clear scaling pattern emerges within model families: larger models and more recent generations tend to occupy the upper-right region of the plot, indicating stronger ability to correctly select the "no right answer" option.
At the same time, most models lie above the diagonal for synonym accuracy and below it for chromosome location accuracy, revealing a systematic gap between the two knowledge types. 
LLMs are generally more successful on synonym questions because synonym-related knowledge is primarily lexical whereas chromosome-related knowledge requires factual and relational understanding beyond linguistic patterns.
This asymmetry suggests that hallucination resistance is uneven across gene attributes.
These results indicate that, although newer and larger LLMs show improved robustness against hallucination, their ability to recognize the absence of valid answers is related to the involved gene information.

% completeness
\subsection{Knowledge Completeness for Multi-Value Gene Attributes}
\label{result:completeness}
\begin{figure}[!ht]
\vspace{-0.4cm}
\centering
\includegraphics[width=0.9\linewidth]{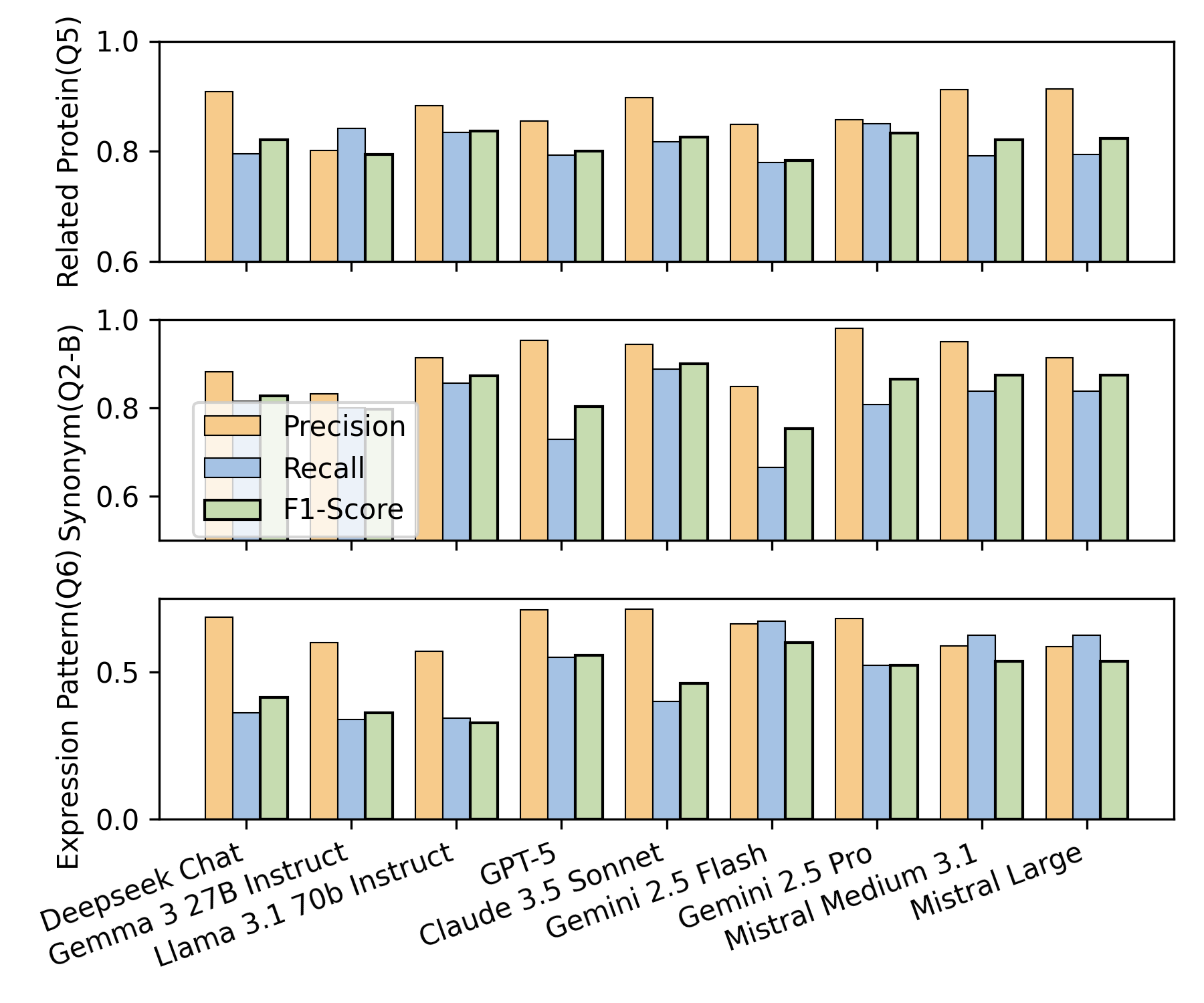}    
\vspace{-0.2cm}
\caption{Completeness evaluation of LLMs. All questions are multiple choice, and the Macro F1-Score is the metric.}
\vspace{-0.2cm}
\label{fig:completeness}
\end{figure}

In this experiment, we answer the question: \textit{Can LLMs provide comprehensive answers for gene attributes that have multiple valid values?}  
Figure~\ref{fig:completeness} reports macro precision, macro recall, and macro F1 scores for three multi-answer tasks: related proteins, synonyms, and expression patterns.
Across all models, precision is almost always higher than recall. 
This pattern indicates that LLMs tend to produce only part of the correct answer set, focusing on items they assign higher confidence to and leaving out other valid answers.
The resulting imbalance suggests that LLM outputs are conservative rather than exhaustive when handling multi-answer genomic questions.
We further observe that macro F1 scores track closely with recall, a trend confirmed by the strong Pearson correlations reported in Table~\ref{correlation}.  
This implies that completeness is primarily constrained by limited recall rather than precision. 
Overall, current LLMs show reasonable accuracy on individual answer components but struggle to produce fully comprehensive responses, revealing a systematic recall deficiency in multi-answer gene tasks.
\begin{table}[!t]
\caption{Pearson correlation coefficients between F1-Score and Precision, and between F1-Score and Recall.}
\resizebox{\columnwidth}{!}{
\begin{tabular}{c|cc|cc|cc}
\bottomrule
\specialrule{0.2pt}{0pt}{0pt}
Knowledge  & \multicolumn{2}{c|}{Related Proteins} & \multicolumn{2}{c|}{Synonym} & \multicolumn{2}{c}{Expression Pattern} \\ \hline
Correlation                                              & Precision           & Recall          & Precision      & Recall      & Precision           & Recall           \\ 
F1-Score                                                 & 0.7526              & \textbf{0.9592}          & 0.8635         & \textbf{0.9712}      & 0.6864              & \textbf{0.9707}           \\ \toprule
\specialrule{0.2pt}{0pt}{0pt}
\end{tabular}}
\vspace{-0.2cm}
\label{correlation}
\end{table}

% literature utilize
\subsection{Influence of Literature Context on Gene-Related Answering}
\label{result:influence_literature}
In this experiment, we answer the question: \textit{Does providing reference abstracts improve LLM performance on gene-related answering tasks?}
We study two tasks where external abstracts may influence the model's behavior: Gene Ontology annotation and functional summary generation.
We summarize metrics across all models under both conditions and visualize their distributions using box plots.
\begin{figure}[htbp]
\vspace{-0.2cm}
    \centering
    \begin{minipage}{0.45\linewidth}
        \centering
        \includegraphics[width=\linewidth]{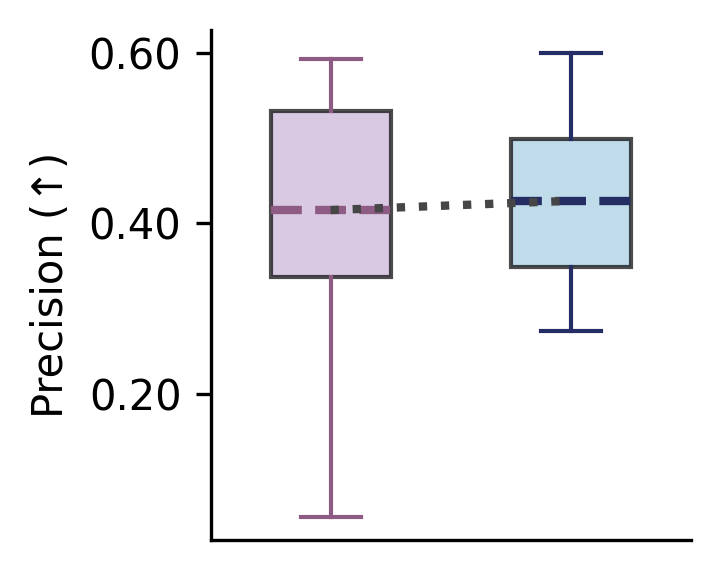}
    \end{minipage}\hfill
    \begin{minipage}{0.45\linewidth}
        \centering
        \includegraphics[width=\linewidth]{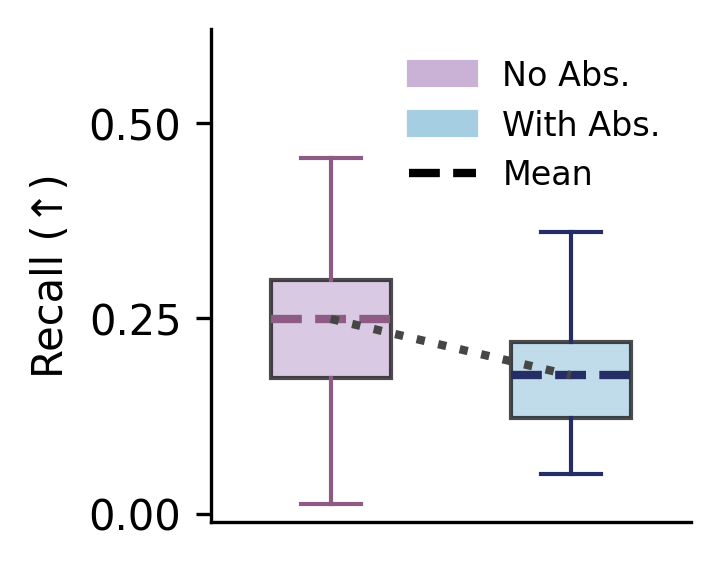}
    \end{minipage}
    \vspace{0.2em}
    \begin{minipage}{0.45\linewidth}
        \centering
        \includegraphics[width=\linewidth]{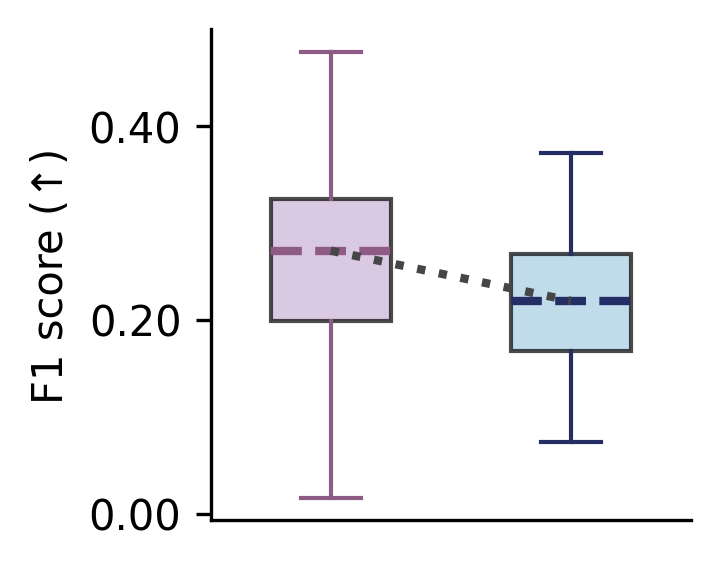}
    \end{minipage}\hfill
    \begin{minipage}{0.45\linewidth}
        \centering
        \includegraphics[width=\linewidth]{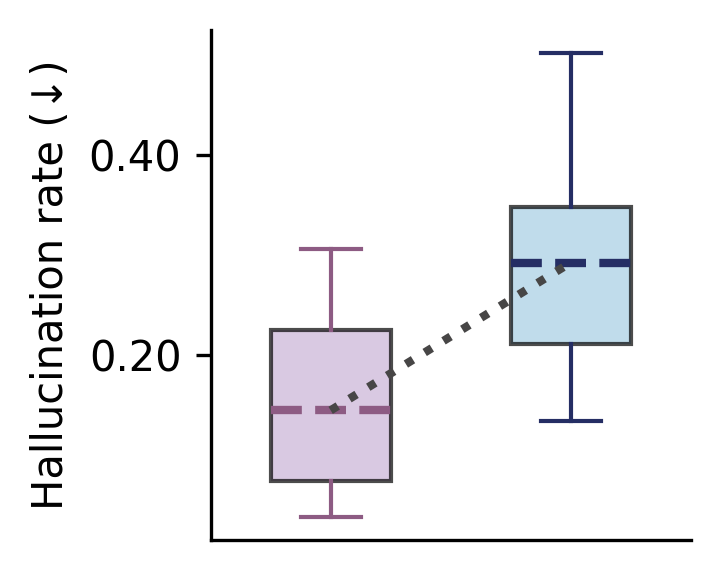}
    \end{minipage}
    \vspace{-0.2cm}
    \caption{Comparison of Gene Ontology answering performance with and without abstracts across all models.}
    \vspace{-0.4cm}
    \label{fig:go_performance}
\end{figure}
\begin{figure}[htbp]
    \vspace{-0.4cm}
    \centering
    \begin{minipage}{0.45\linewidth}
        \centering
        \includegraphics[width=\linewidth]{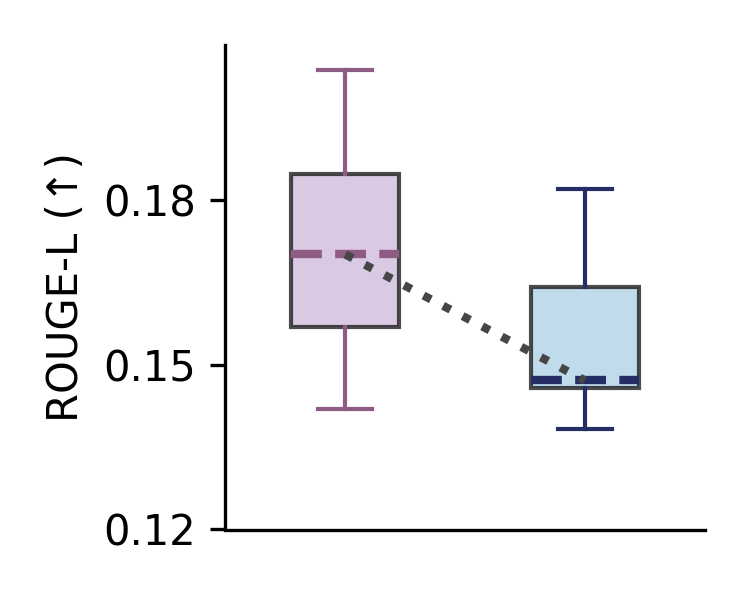}
    \end{minipage}\hfill
    \begin{minipage}{0.45\linewidth}
        \centering
        \includegraphics[width=\linewidth]{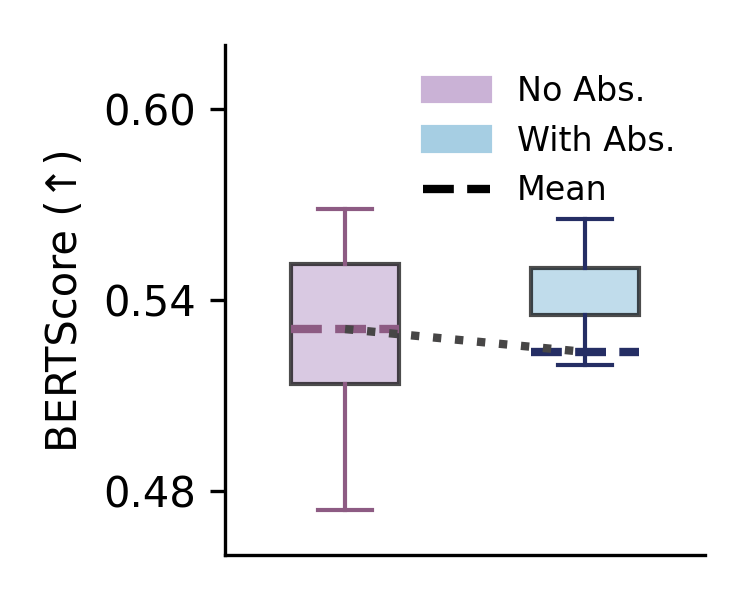}
    \end{minipage}
    \vspace{0.2em}
    \begin{minipage}{0.45\linewidth}
        \centering
        \includegraphics[width=\linewidth]{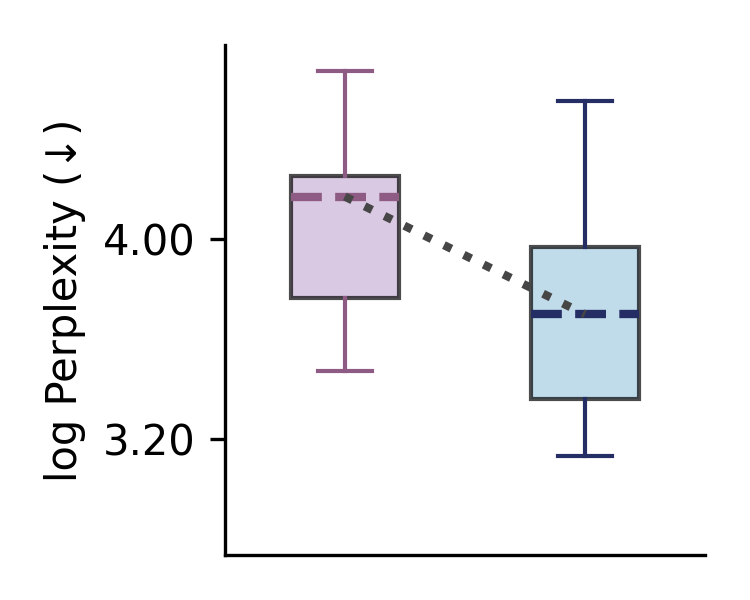}
    \end{minipage}\hfill
    \begin{minipage}{0.45\linewidth}
        \centering
        \includegraphics[width=\linewidth]{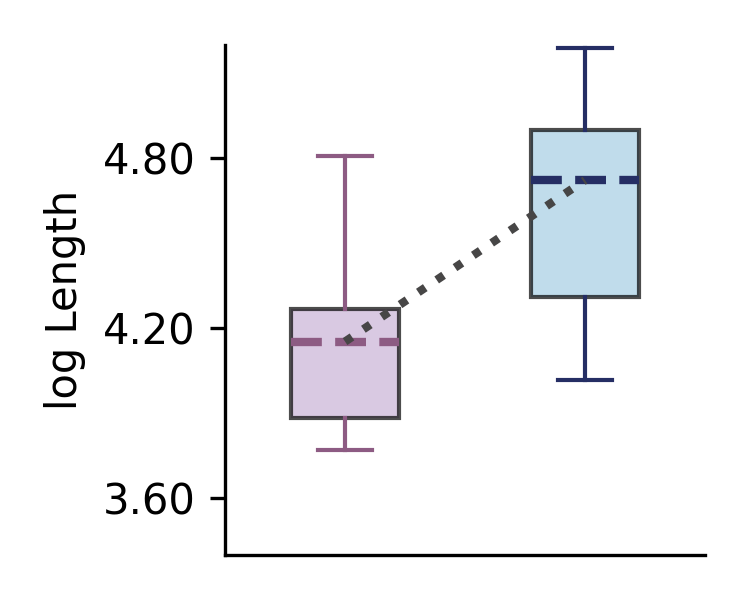}
    \end{minipage}
    \vspace{-0.4cm}
    \caption{Comparison of functional summary answering performance with and without abstracts across all models.}
    \label{fig:summary_performance}
    \vspace{-0.2in}
\end{figure}

For Gene Ontology questions, Figure~\ref{fig:go_performance} shows that supplying abstracts does not lead to consistent improvements across models.
The precision distribution becomes more concentrated when abstracts are provided, suggesting that models tend to converge toward similar predictions.
However, the mean precision remains comparable to the no-abstract condition, indicating no clear gain in correctness.
Recall and F1 shift downward, and hallucination rates increase, showing that models output fewer valid GO terms and generate more unsupported ones.
These outcomes suggest that LLMs may struggle to map unstructured narrative information from abstracts to the controlled vocabulary of the ontology.
Instead of extracting standardized terms, models may rely on salient phrases or co-occurring concepts in the abstracts, which shifts predictions away from the gold annotations.

For functional summary generation, the effect of abstracts is also mixed.
As shown in Figure~\ref{fig:summary_performance}, ROUGE-L exhibits a slight downward shift, while BERTScore remains broadly similar across the two conditions.
Perplexity decreases and its distribution narrows, indicating that summaries become more fluent when abstracts are provided.
Meanwhile, the outputs become longer as models incorporate details from the abstracts into their generated summaries.
These shifts show that reference-conditioned summaries are more verbose and textually smooth but diverge from the concise style of curated functional summaries.

Overall, reference abstracts influence model outputs in systematic ways but do not consistently enhance task performance.
The additional context guides models toward patterns present in the abstracts, yet their ability to transform this information into structured or distilled gene-level answers remains limited.
Thus, literature context affects model behavior, but its utility for improving gene-level answering remains limited.

\begin{table*}[!ht]
\caption{The overall results of LLMs. Higher values indicate better performance. The best score for each question type is marked in "\textbf{bold}", and the second-best score is \underline{underlined}. Overall scores are computed by equally averaging across tasks.}
\resizebox{0.97\textwidth}{!}{
\begin{tabular}{lccccccccc}
\specialrule{1.2pt}{0pt}{0pt} % 顶部双线（加粗）
& \makecell{Gene Full\\Name} & Synonyms & \makecell{Chromosome \\Location} & \makecell{Gene\\Type} & \makecell{Related\\Proteins} & \makecell{Expression\\Pattern} & \makecell{Gene\\Ontology} & \makecell{Functional\\Summary} & Overall \\ \hline
\multicolumn{10}{c}{Close Source Models} \\ \hline % <--- 正确的居中标题
GPT-5~\cite{singh2025openai} & \textbf{67.68} & 83.91 & \underline{53.53} & \underline{73.93} & 82.39 & \underline{55.02} & \textbf{43.17} & 35.70 & \textbf{61.92} \\
Claude 3.5 Sonnet~\cite{anthropic2024claude} & 67.14 & \underline{87.86} & 47.86 & \textbf{74.23} & 82.66 & 43.35 & 34.58 & 37.13 & \underline{59.35} \\
Mistral Medium 3.1~\cite{mistralai2025medium31} & 66.67 & 84.31 & 48.77 & 71.49 & 82.18 & 50.33 & 34.46 & 35.17 & 59.17 \\
Gemini 2.5 Pro~\cite{comanici2025gemini} & 64.68 & 82.48 & 44.92 & 64.80 & 83.38 & 53.58 & 35.51 & 36.19 & 58.19 \\
Claude 4.5 Sonnet & 62.74 & 75.02 & 38.14 & 73.22 & \textbf{84.80} & 45.73 & \underline{39.23} & 37.43 & 57.04 \\
GPT-4o~\cite{hurst2024gpt4o} & 65.77 & 84.55 & \textbf{56.23} & 69.70 & 82.28 & 35.47 & 17.57 & 37.36 & 56.12 \\
GPT-4 Turbo~\cite{achiam2023gpt4} & \underline{67.38} & 75.84 & 46.98 & 61.01 & 83.01 & 32.99 & 31.91 & \textbf{37.66} & 54.60 \\
Gemini 2.5 Flash~\cite{comanici2025gemini} & 63.18 & 69.25 & 42.80 & 64.05 & 77.39 & \textbf{56.02} & 26.33 & 34.22 & 54.15 \\
GPT-3.5 Turbo~\cite{ouyang2022gpt3} & 64.58 & 60.61 & 14.02 & 63.31 & 80.38 & 22.70 & 25.40 & 35.33 & 45.79 \\ \hline
\multicolumn{10}{c}{Open Source Models: Large Scale} \\ \hline % <--- 正确的居中标题
DeepSeek Chat-671B~\cite{liu2024deepseek} & 65.83 & \textbf{89.59} & 39.60 & 67.92 & 82.10 & 40.94 & 28.83 & 36.65 & 56.43 \\
Mistral Large-123B~\cite{mistral2024mistrallarge} & 66.73 & 84.25 & 48.61 & 71.55 & 82.36 & 24.82 & 32.33 & 36.51 & 55.89 \\
LLaMA 3.1-405B~\cite{grattafiori2024llama3} & 64.05 & 82.54 & 36.27 & 60.30 & \underline{83.66} & 22.73 & 23.00 & \underline{37.55} & 51.26 \\
Mixtral 8x22B (141B)~\cite{jiang2024mixtral} & 65.12 & 70.15 & 27.34 & 57.98 & 81.06 & 39.29 & 24.24 & 35.71 & 50.11 \\ \hline
\multicolumn{10}{c}{Open Source Models: Medium Scale} \\ \hline % <--- 正确的居中标题
Qwen 2.5-72B~\cite{qwen2.5} & 66.73 & 83.44 & 40.87 & 65.18 & 77.58 & 30.26 & 31.48 & 36.64 & 54.02 \\
Mistral Small 3.1-24B~\cite{mistral2025small31} & 65.06 & 74.97 & 34.98 & 57.38 & 77.10 & 37.46 & 19.00 & 35.86 & 50.23 \\
LLaMA 3.1-70B~\cite{grattafiori2024llama3}  & 64.29 & 54.65 & 23.17 & 72.26 & 83.69 & 33.60 & 28.79 & 36.00 & 49.56 \\
Gemma 3 27B~\cite{gemmateam2025gemma3technicalreport} & 62.56 & 45.15 & 14.60 & 63.93 & 79.40 & 34.58 & - & 34.02 & 47.75 \\ 
*MedGemma-27B Text-it~\cite{sellergren2025medgemma} & 55.66 & 49.67 & 17.98 & 53.09 & 67.78 & 38.28 & 16.26 & 31.81 & 41.32 \\
*PMC-LLaMA-13B~\cite{wu2024pmc} & 57.32 & 43.85 & 20.68 & 29.23 & 46.73 & 7.81 & 2.77 & 16.71 & 28.14 \\\hline
\multicolumn{10}{c}{Open Source Models: Small Scale} \\ \hline % <--- 正确的居中标题
Ministral-8B~\cite{mistral2024ministral8b} & 62.32 & 81.21 & 17.42 & 57.86 & 72.42 & 24.12 & 16.27 & 36.25 & 45.98 \\
Qwen 2.5-7B~\cite{qwen2.5} & 64.17 & 76.33 & 27.66 & 55.95 & 68.56 & 24.35 & 11.22 & 29.61 & 44.73 \\
Gemma 3-4B~\cite{gemmateam2025gemma3technicalreport} & 53.15 & 38.41 & 14.09 & 56.90 & 72.16 & 19.87 & 11.07 & 32.82 & 37.31 \\
Mistral-7B~\cite{jiang2023mistral7b} & 50.71 & 44.65 & 17.18 & 45.12 & 78.47 & 7.55 & 2.84 & 25.54 & 34.01 \\
LLaMA 3.1-8B~\cite{grattafiori2024llama3} & 62.56 & 38.12 & 14.56 & 54.47 & 77.15 & 30.59 & - & 34.54 & 44.57 \\
*BioMistral-7B~\cite{labrak:hal-04621178} & 32.38 & 48.81 & 15.80 & 29.64 & 73.45 & 21.33 & 9.77 & 37.47 & 33.58 \\
*Biomedical-LLaMA-3-8B~\cite{Bio-Medical-Llama-3-8B} & 38.75 & 32.61 & 15.16 & 29.17 & 67.23 & 45.34 & 7.58 & 29.63 & 33.18 \\
*MedAlpaca-7B~\cite{pal2022medmcqa} & 45.00 & 29.93 & 15.44 & 27.74 & 43.43 & 0.00 & - & 32.03 & 27.65 \\ \hline
\specialrule{1.2pt}{0pt}{0pt}
\end{tabular}
}
\label{tab:main}
\end{table*}

\subsection{Top Model Comparison Across Evaluation Dimensions}
\label{result:top_models}
\begin{figure}[ht]
\centering
\includegraphics[width=0.9\linewidth]{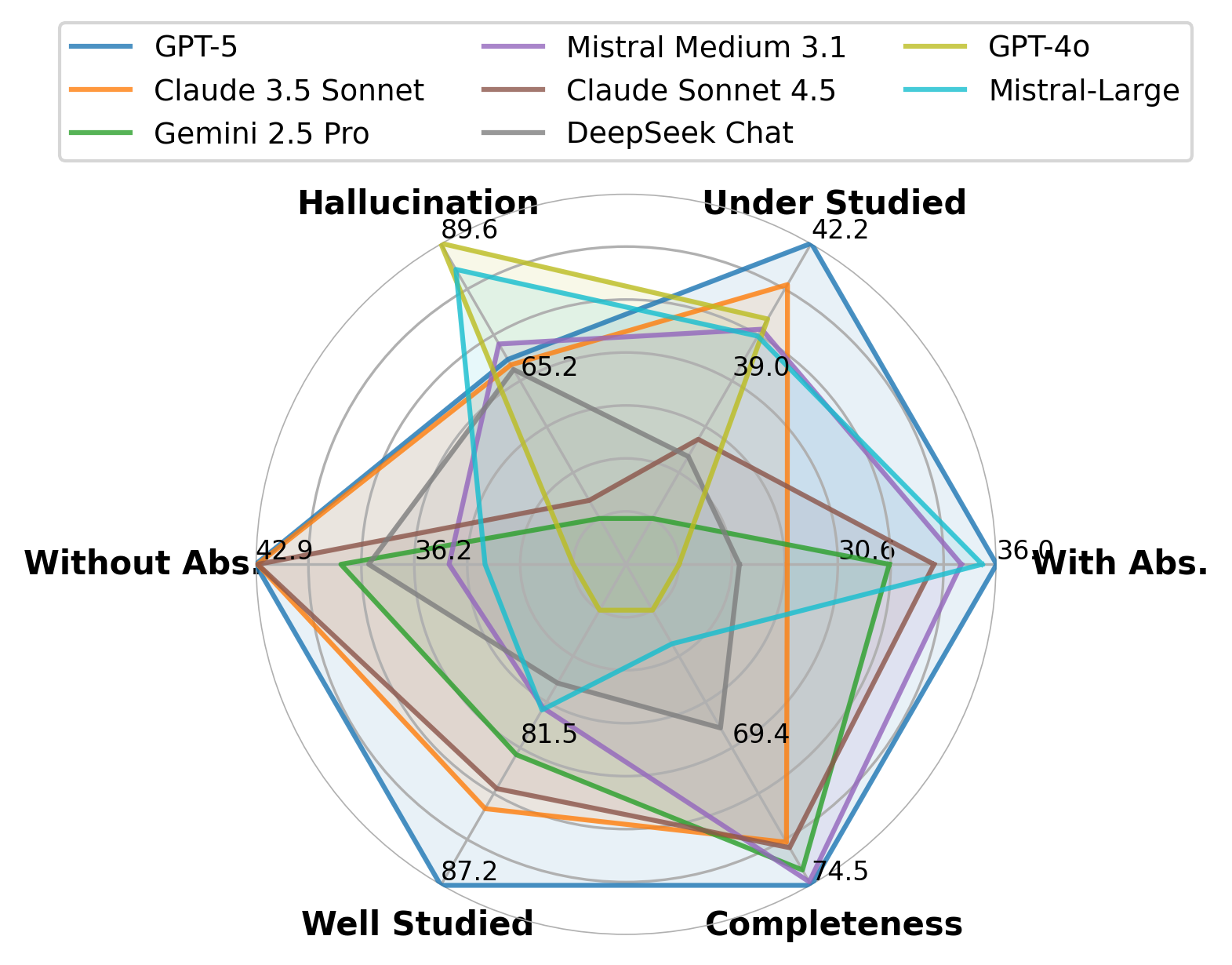}
    \caption{Performance profiles of top-performing models across different evaluation dimensions.}
    \vspace{-0.2in}
    \label{fig:top_performance}
\end{figure}
In this section, we aim to study: \textit{What characteristics distinguish the top-performing models?} 
We compare top eight LLMs across all evaluation dimensions.
As shown in Figure~\ref{fig:top_performance}, these models exhibit distinct and non-overlapping capability profiles. 
Strong performance in one dimension does not reliably predict strength in another. 
For instance, GPT-4o shows the highest resistance to hallucination but achieves only moderate completeness. 
Such divergence indicates that the evaluation dimensions capture largely independent facets of gene-related understanding.
Overall, the results show that top-performing models achieve their effectiveness through different combinations of strengths rather than uniform superiority. 
Therefore, characterizing LLM capabilities in gene-centered tasks requires a multi-dimensional evaluation framework instead of relying on any single metric or task category.

\begin{table*}[!t]
\centering
\caption{Case study on the SHOC2 functional summary task, comparing model performance with and without literature context across multiple evaluation metrics.}
\label{tab:shoc2_case_study}
\scriptsize
\resizebox{\textwidth}{!}{
\begin{tabular}{lcccccccc}
\toprule
\textbf{Model} 
& \textbf{ROUGE-L} 
& \textbf{ROUGE-L} 
& \textbf{BERTScore} 
& \textbf{BERTScore} 
& \textbf{PPL} 
& \textbf{PPL} 
& \textbf{Length} 
& \textbf{Length} \\
& \textbf{No Ref} 
& \textbf{With Ref ($\Delta$)} 
& \textbf{No Ref} 
& \textbf{With Ref ($\Delta$)} 
& \textbf{No Ref} 
& \textbf{With Ref ($\Delta$)} 
& \textbf{No Ref} 
& \textbf{With Ref} \\
\midrule
GPT-4o                & 0.225 & 0.185 (-17.8\%) & 0.592 & 0.580 (-2.0\%)  & 31.9 & 21.3 (-33.2\%) & 51  & 137 \\
GPT-4 Turbo           & 0.214 & 0.140 (-34.6\%) & 0.583 & 0.531 (-8.9\%)  & 33.5 & 27.0 (-19.4\%) & 44  & 141 \\
DeepSeek Chat-671B    & 0.306 & 0.144 (-52.9\%) & 0.663 & 0.580 (-12.5\%) & 47.6 & 38.7 (-18.7\%) & 62  & 75  \\
Claude 4.5 Sonnet     & 0.370 & 0.204 (-44.9\%) & 0.698 & 0.615 (-11.9\%) & 39.9 & 28.4 (-28.8\%) & 126 & 136 \\
Gemini 2.5 Flash      & 0.198 & 0.171 (-13.6\%) & 0.600 & 0.567 (-5.5\%)  & 65.1 & 36.2 (-44.4\%) & 49  & 72  \\
Biomedical-LLaMA-3-8B & 0.247 & 0.218 (-11.7\%) & 0.583 & 0.598 (+2.6\%)  & 44.1 & 27.3 (-38.1\%) & 31  & 64  \\
\bottomrule
\end{tabular}
}
\end{table*}
\subsection{Overall Performance across LLMs}
\label{result:overall}
In this section, we investigate the question: \textit{How do different LLMs perform on human gene-knowledge tasks?} 
Table~\ref{tab:main} presents the overall performance of LLMs across the eight curated question types. 
Closed-source models generally achieve higher accuracy than open-source models, reflecting the advantages conferred by more extensive optimization in model architecture, training data quality, and instruction alignment. 
We also observe that domain-specific biomedical models do not exhibit a consistent performance advantage. 
A plausible reason is that, although these models are trained on biomedical corpora such as PubMed papers or clinical narratives, such text sources contain limited coverage of the structured and fine-grained gene attributes. 
As a result, these models receive little gene-level supervision and their learned domain knowledge does not directly translate to the factual precision needed for gene-centered questions for our benchmark. 
These findings indicate that closed-source models demonstrate overall superiority, while current domain-specific models provide only limited benefits due to limited gene-specific supervision and insufficient alignment for fine-grained gene-level understanding.

\section{Case Study}
In this section, we present a case study on the SHOC2 functional summary task. 
This example illustrates how abstracts alter model outputs at the content level. 
As shown in Table~\ref{tab:shoc2_case_study}, adding abstracts yields more fluent outputs, with perplexity decreasing by 18\%--44\%, and substantially longer responses (e.g., GPT-4o: 51->137; GPT-4-turbo: 44->141), while ROUGE decreases and BERTScore changes more moderately. 
This pattern suggests that additional information shifts the output content distribution.
Table~\ref{tab:shoc2_case_study_text} presents responses from GPT-4o and DeepSeek-Chat.
\begin{table}[ht]
\centering
\vspace{-0.1in}
\caption{Case study of the SHOC2 functional summary task. 
\textbf{Bold} indicates content aligned with the ground-truth reference, while \uline{underlined text} denotes additional information introduced from literature context.}
\label{tab:shoc2_case_study_text}
\footnotesize
\setlength{\tabcolsep}{3pt}
\renewcommand{\arraystretch}{1.15}
\begin{tabularx}{\columnwidth}{p{0.23\columnwidth}X}
\toprule
\textbf{Source} & \textbf{Functional summary} \\
\midrule

Ground truth 
& This gene encodes a protein that consists almost entirely of 
\textbf{leucine-rich repeats}, a domain implicated in 
\textbf{protein-protein interactions}. The protein may function as a 
\textbf{scaffold linking RAS to downstream signal transducers in the RAS/ERK MAP kinase signaling cascade}. 
Mutations in this gene have been associated with Noonan-like syndrome with loose anagen hair. \\

\midrule

GPT-4o w/o ref. 
& SHOC2 is a \textbf{scaffold protein} that facilitates the activation of the 
\textbf{RAS-RAF-MEK-ERK signaling pathway} by promoting the dephosphorylation of RAF1, 
thereby enhancing its \textbf{interaction with RAS} and subsequent downstream signaling. \\

GPT-4o w/ ref. 
& SHOC2 is a protein that functions as a \textbf{scaffold} in the 
\textbf{Ras-MAPK signaling pathway}, facilitating the 
\textbf{interaction between Ras and Raf}. It forms a complex with the catalytic subunit of 
\uline{protein phosphatase 1 (PP1c)} to dephosphorylate the 
\uline{S259 inhibitory site of Raf proteins}, thereby stimulating Raf activity. 
SHOC2 is essential for the activation of the \textbf{MAPK pathway} by 
\uline{growth factors} and is implicated in mediating 
\uline{FGF receptor signaling}. \uline{In tumor cells, it may also be relevant for cancer treatment.} \\

\midrule

DeepSeek-Chat w/o ref. 
& SHOC2 is a \textbf{leucine-rich repeat-containing protein} that functions as a 
\textbf{scaffold} to positively regulate the 
\textbf{RAS-RAF-MEK-ERK signaling pathway} by promoting the formation of a ternary complex with 
\textbf{RAS} and RAF, facilitating RAF activation and 
\textbf{downstream MAPK signaling}. \\

DeepSeek-Chat w/ ref. 
& SHOC2 is a \textbf{scaffold protein} that positively regulates 
\textbf{Ras-MAP kinase signaling} by forming a complex with 
\textbf{Ras} and Raf, \textbf{facilitating their interaction} and enhancing Raf activation through dephosphorylation of its inhibitory site by 
\uline{protein phosphatase 1 (PP1c)}. It is essential for 
\uline{growth factor}-induced MAPK pathway activation and is implicated in 
\uline{cancer cell signaling}. \\
\bottomrule
\end{tabularx}
\end{table}
For GPT-4o, the no-reference response gives a concise summary focused on the core pathway role. 
With references, the model adds details such as PP1c, growth factor signaling, and tumor relevance that are grounded in the abstracts. 
These additions enrich the biological context but deviate from the ground-truth reference, reducing lexical and semantic similarity.
A similar pattern appears for DeepSeek-Chat. 
Without references, it produces a relatively complete and structured summary aligned with the ground-truth functional description. 
With references, it incorporates abstract-grounded context, including PP1c-mediated dephosphorylation, growth factor signaling, and cancer-related processes. 
While biologically informative, these additions go beyond the curated reference and again reduce similarity scores.
Overall, the shifts are not merely superficial responses to added text, but reflect systematic incorporation of context-specific biological information.

\section{Conclusions}
% This work introduces \bench{}, a large-scale gene-centric benchmark for evaluating LLMs' gene-to-function reasoning.
% Built on an authentic source, \bench{} assesses model behavior across four biologically critical perspectives: research-attention sensitivity, hallucination tendency, knowledge completeness, and literature influence.
% Toward evaluations, we identify substantial heterogeneity and a consistent reliability gap: strong biomedical QA performance does not necessarily translate into faithful gene-level functional interpretation.
% Crucially, gene-level failures are not random but follow systematic degradation patterns, with reliability predictably decreasing for low-attention genes, sparse evidence settings, multi-answer queries, and shifts in reference context.
% We further observe that domain-specialized models do not consistently outperform rapidly advancing general-purpose LLMs, which often achieve comparable or superior reliability across multiple evaluation perspectives.
% Overall, \bench{} provides a principled and scalable foundation for characterizing gene-level understanding in LLMs, offering actionable insights for model selection and development in biologically grounded interpretation settings.
This work introduces \bench{}, a large-scale gene-centric benchmark for evaluating LLMs' gene-to-function reasoning.
Built on authoritative biological sources, \bench{} assesses model behavior across research-attention sensitivity, hallucination tendency, knowledge completeness, and literature influence.
Our evaluations reveal substantial heterogeneity and a persistent reliability gap: strong biomedical QA performance does not necessarily translate into faithful gene-level functional interpretation.
Gene-level failures follow systematic degradation patterns, especially for low-attention genes, sparse evidence settings, multi-answer queries, and shifts in reference context.
We further find that domain-specialized models do not consistently outperform general-purpose LLMs.
Overall, \bench{} provides a scalable foundation for characterizing gene-level understanding in LLMs and supporting model selection in biologically grounded interpretation settings.

% \begin{displaymath}
% \end{displaymath}

\balance

\begin{acks}
This work is partially supported by the National Natural Science Foundation of China (No.62506351), the Strategic Priority Research Program of the Chinese Academy of Sciences (No.XDB1350102), the National Natural Science Foundation of China (92470204, 62506352), and the Beijing Natural Science Foundation (No.4254089).
The development of this project was supported by the Scihorizon platform.
\end{acks}

\clearpage
\bibliographystyle{ACM-Reference-Format}
\bibliography{aa_sample-base}

\appendix
% \clearpage
% appendix 2

\appendix
\setcounter{table}{0}
\setcounter{figure}{0}
\renewcommand{\thetable}{S\arabic{table}}
\renewcommand{\thefigure}{S\arabic{figure}}

\section{Benchmark Construction and Statistics}
\label{appendix:benchmark-details}

\begin{table}[H]
\centering
\caption{Overview of benchmark composition across evaluation perspectives, question types, and scale.}
\resizebox{\linewidth}{!}{
\begin{tabular}{ccc}
\specialrule{1.2pt}{0pt}{0pt}
Evaluation Perspective & Question Type                                                              & Task  Scale                                                                          \\ \hline
Research Attention     & Single Choice                                                              & \begin{tabular}[c]{@{}c@{}}High Attention:136{,}676\\ Low Attention:290{,}370\end{tabular} \\ \hline
Hallucination Tendency & \begin{tabular}[c]{@{}c@{}}Single Choice\\ Templated Answer\end{tabular}   & 36{,}444                                                                                \\ \hline
Answer Completness     & \begin{tabular}[c]{@{}c@{}}Multiple Choice\\ Templated Answer\end{tabular} & 61{,}726                                                                                \\ \hline
Literature Influence   & \begin{tabular}[c]{@{}c@{}}Templated Answer\\ Open Answer\end{tabular}     & 40{,}452                                                                                \\ \specialrule{1.2pt}{0pt}{0pt}
\end{tabular}
}
\label{tab:benchmark}
\end{table} 
% \FloatBarrier
% \input{table/benchmark_details_knowledge_perspective}
This section provides detailed statistics of the benchmark.
Table~\ref{tab:benchmark} summarizes the number of questions across four evaluation perspectives.
In total, the benchmark consists of over 540K questions derived from more than 190K human genes.
The questions span single-choice, multiple-choice, templated generation, and open-form answering tasks.
% \FloatBarrier

% \input{table/compare_models}
% \FloatBarrier

% We evaluate 27 LLMs, including general-purpose and biomedical domain-specialized models.
% Closed-source models are accessed via APIs using default inference settings.
% Open-source and domain-specific models are deployed locally following their released configurations.
% A full list of evaluated models and their versions is provided in Table~\ref{tab:models}.

\section{Evaluation Protocol and Implementation Details}
\label{appendix:evaluation-protocol}
\begin{figure*}[!t]
    \centering
    \includegraphics[width=0.45\linewidth]{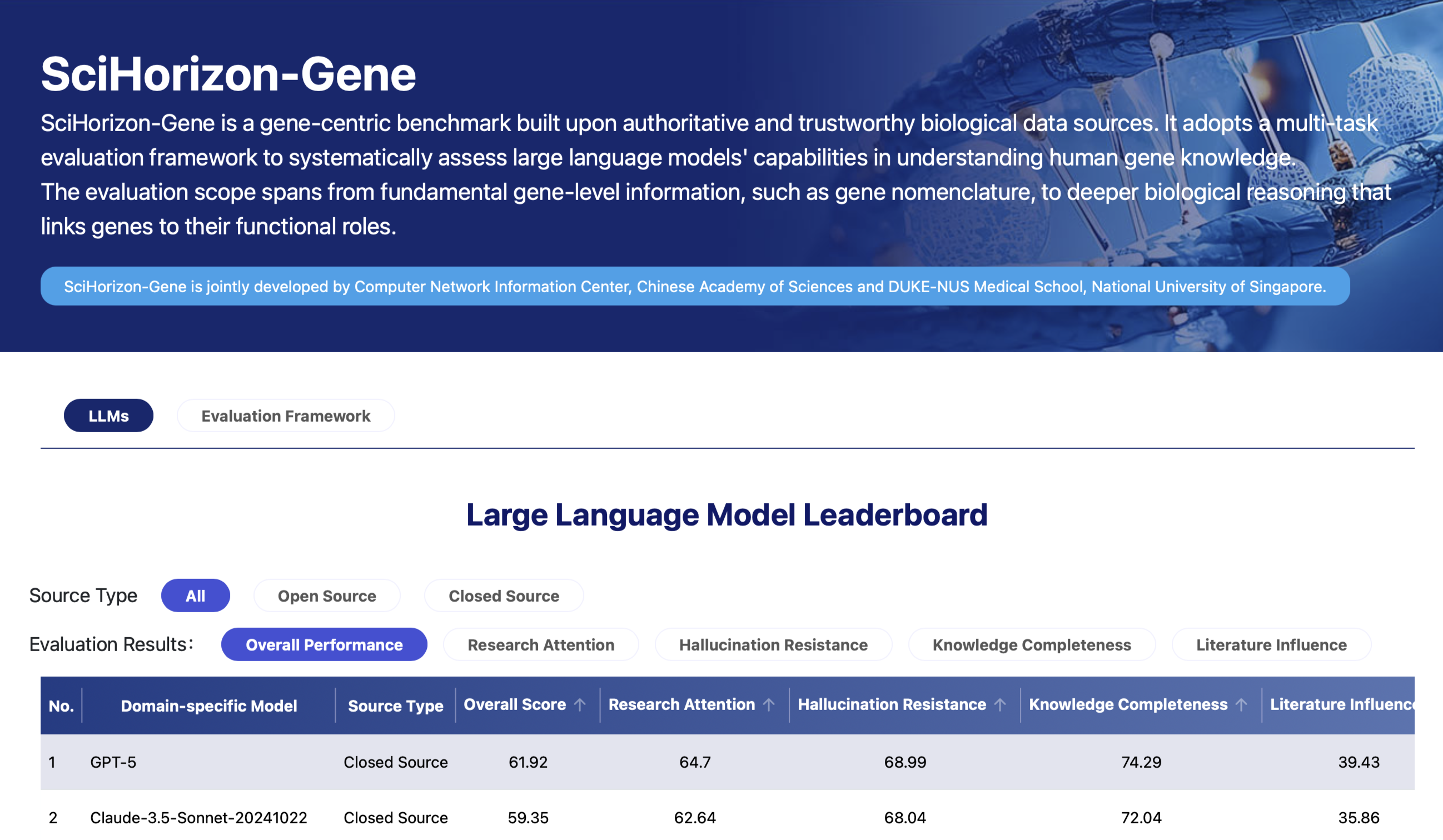}
    \hfill
    \includegraphics[width=0.45\linewidth]{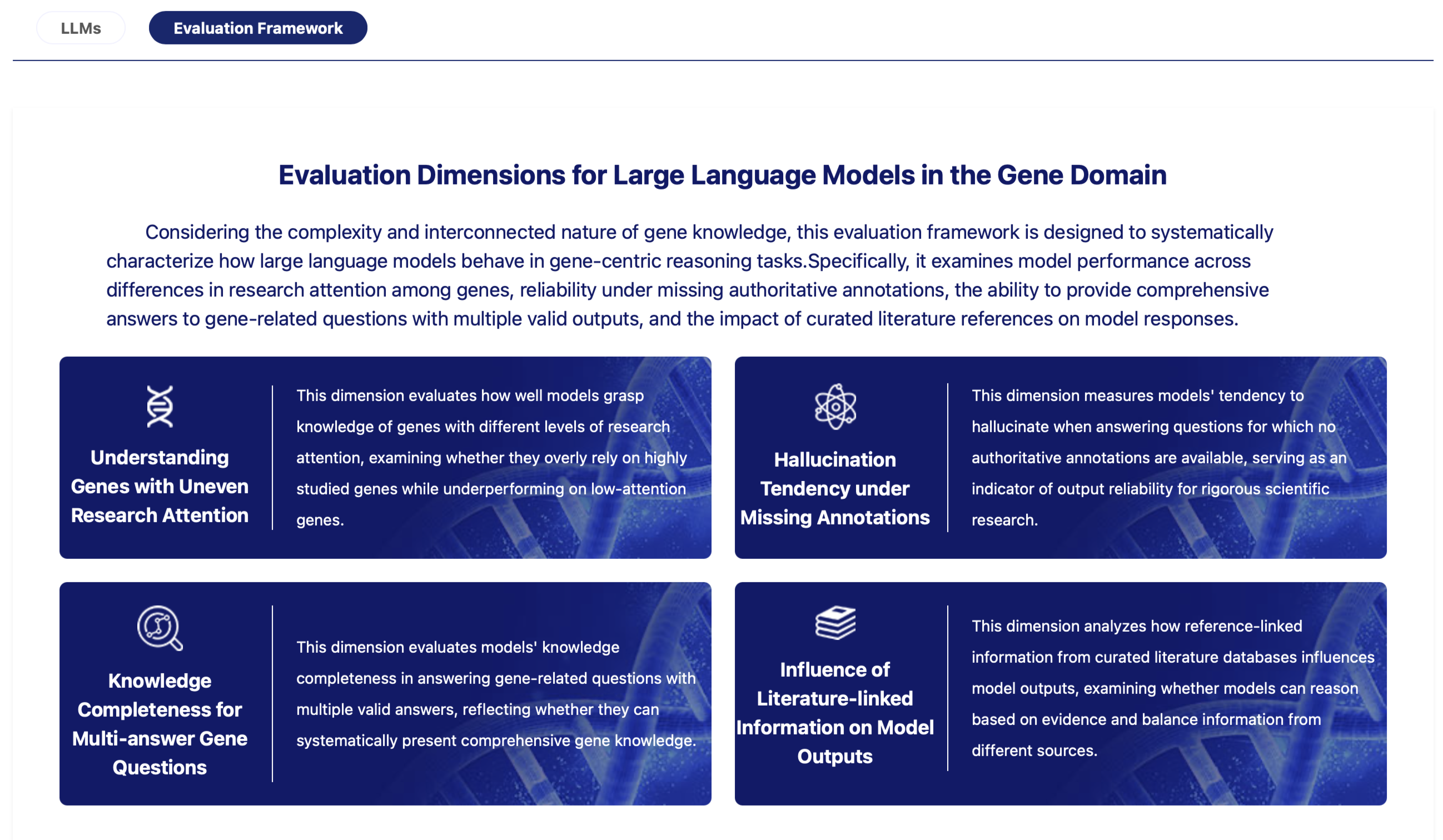}
    \caption{
    Left: The SciHorizon-Gene online platform for presenting evaluation results, supporting efficient filtering and ranking of models.
    Right: A systematic overview of the SciHorizon-Gene evaluation framework, illustrating the organization of evaluation perspectives.
    }
    \label{fig:online}
\end{figure*}
To reduce positional bias in choice evaluation~\cite{wang2024large}, we employ a cyclic permutation strategy that systematically rotates the correct option across all possible positions.
Positional bias refers to LLMs favoring certain option positions (e.g., earlier or later choices) regardless of content, which may distort evaluation results.
Specifically, for each choice question, we generate a set of permuted variants in which the correct answer appears once in every candidate position. 
The model is evaluated on all permuted versions, and the final results are aggregated across permutations to obtain a position-invariant assessment.
A subset of 2,710 representative questions is sampled to cover all biological scenarios and evaluation perspectives. 
After applying the cyclic permutation to mitigate positional bias, this results in 9,010 evaluation instances per model. 
All queries in this set are conducted in a zero-shot setting, with the temperature fixed to 0 to ensure reproducibility.
Domain-specific models are evaluated following their official usage instructions and recommended inference settings.

\section{Evaluation Metrics}
\label{appendix:metrics}
% To enable automated and fine-grained assessment of LLM performance, we design task-specific evaluation metrics for five categories of questions. 
% Each question type reflects a distinct aspect of gene-related knowledge and requires tailored evaluation procedures.
We design task-specific evaluation metrics and each question type reflects a distinct aspect of gene-related knowledge.

\textbf{Single-Choice Questions}
% We use single-choice questions to evaluate LLMs' understanding of genes with varying research attention and their ability to correctly identify cases with no valid answer.
% The evaluation metric is accuracy, defined as:
% \begin{equation}
% \mathrm{Accuracy} = 
% \frac{\text{\# correctly answered questions}}{\text{\# total questions}}
% \end{equation}
% Higher accuracy indicates better recognition of correct gene-specific knowledge and stronger hallucination resistance against selecting invalid options.
% The overall performance is reported using averaged accuracy across multiple evaluation.
\sloppy
Single-choice questions evaluate LLMs' gene-specific understanding under varying research attention and their ability to recognize cases with no valid answer.
Higher accuracy indicates stronger knowledge recognition and hallucination resistance.

\textbf{Multiple-Choice Questions}
% Multiple-choice questions assess the LLMs' completeness in understanding gene-related knowledge.
% Let $R$ denote the reference set of correct options and $P$ the set of predicted options. 
% We compute precision, recall, and F1-score for each question:
% \begin{equation}
% \mathrm{Precision} = \frac{|P \cap R|}{|P|} \quad
% \mathrm{Recall} = \frac{|P \cap R|}{|R|},
% \end{equation}
% \begin{equation}
% \mathrm{F1} =
% \frac{2 \times \mathrm{Precision} \times \mathrm{Recall}}
% {\mathrm{Precision} + \mathrm{Recall}}
% \end{equation}
% The overall performance is reported using macro-averaged F1.
Multiple-choice questions assess the completeness of LLMs in understanding gene-related knowledge.
Let $R$ and $P$ denote the reference and predicted option sets, respectively.
For each question, we compute precision $=|P\cap R|/|P|$, recall $=|P\cap R|/|R|$, and
$\mathrm{F1}=2\,\mathrm{Precision}\cdot\mathrm{Recall}/(\mathrm{Precision}+\mathrm{Recall})$.
Overall performance is reported using macro-averaged F1.

\textbf{Expression Pattern}
% For gene expression pattern questions, each reference answer is represented as a structured dictionary containing an expression category (\texttt{Category}) and a list of expressed tissues (\texttt{Tissue}).  
% The expression category is treated as a scalar label and is scored as 1 for an exact match and 0 otherwise. 
% The tissue list is treated as a multi-label set, for which an F1-score is computed.  
% The final expression score is defined as:
% \begin{equation}
% \mathrm{Score}_{\mathrm{Expr}}
% = \alpha \cdot \mathrm{CategoryMatch}
% + \beta \cdot \mathrm{F1}_{\mathrm{Tissue}}
% \end{equation}
% where both components are given equal weight, i.e., $\alpha = \beta = 0.5$.
For gene expression pattern questions, each reference answer contains an expression category and expressed tissues. 
We score the category by exact match and the tissue list by multi-label F1, and define:
$
\mathrm{Score}_{\mathrm{Expr}}
= \alpha \cdot \mathrm{CategoryMatch}
+ \beta \cdot \mathrm{F1}_{\mathrm{Tissue}},
$
where $\alpha=\beta=0.5$.

\textbf{Gene Ontology Annotation}
% We follow the procedure widely adopted in the \textit{Critical Assessment of Function Annotation} (CAFA)~\cite{jiang2016expanded} for Gene Ontology annotation evaluation.
% We introduce minor adaptations for language model outputs. 
% All predicted GO descriptions are first normalized and mapped to valid GO identifiers using a dictionary constructed from official terms and their synonyms (i.e., from \texttt{go-basic.obo}).
% Unmappable predictions are excluded from the evaluation and tracked separately as hallucinated outputs. 
% For each gene, we then compute the GO closures of both the reference annotations and the model predictions, and derive precision, recall, and F1-score from the resulting sets.
We follow the CAFA protocol~\cite{jiang2016expanded} for Gene Ontology annotation evaluation, with minor adaptations for LLM outputs.
Predicted GO descriptions are normalized and mapped to valid GO identifiers using official terms from \texttt{go-basic.obo}.
Unmappable predictions are excluded from scoring and counted as hallucinated outputs.
For each gene, we compute GO closures for both references and predictions, then derive precision, recall, and F1 from the resulting sets.

\textbf{Functional Summary}
% For functional summary generation, we regard the NCBI functional description of each gene as the reference text. 
% The evaluation integrates three complementary dimensions: ROUGE-L~\cite{lin2004rouge} for lexical overlap, BERTScore for semantic similarity~\cite{zhang2019bertscore}, and perplexity (PPL) as a measure of fluency~\cite{manning1999foundations}.  
% We apply following transformation to normalize perplexity into a fluency score, so that higher score means better performance:
% \begin{equation}
% \mathrm{Fluency} =
% \frac{1}{1 + \log_{10}(\mathrm{PPL})}.
% \end{equation}
% The final summary score is computed as the arithmetic mean of the three components:
% \begin{equation}
% \mathrm{Score}_{\mathrm{Func}}
% = \frac{1}{3}
% \left(
% \mathrm{ROUGE\mbox{-}L}
% + \mathrm{BERTScore}
% + \mathrm{Fluency}
% \right).
% \end{equation}
For functional summary generation, we use the NCBI functional description as the reference text and evaluate outputs with ROUGE-L~\cite{lin2004rouge}, BERTScore~\cite{zhang2019bertscore}, and a PPL-based fluency score~\cite{manning1999foundations}:
\begin{equation}
\mathrm{Fluency} =
\frac{1}{1 + \log_{10}(\mathrm{PPL})}.
\end{equation}
The final score is the mean of the three components:
\begin{equation}
\mathrm{Score}_{\mathrm{Func}}
= \frac{\mathrm{ROUGE\mbox{-}L}+\mathrm{BERTScore}+\mathrm{Fluency}}{3}.
\end{equation}

% \section{Detailed Results}
% \label{appendix:detailed_results}
% \input{table/dimension_performance_1}
% \input{table/dimension_performance_2}
% Table~\ref{tab:research_attention&hallucination} and Table~\ref{tab:knowledge_completeness&literature} report the detailed performance of all evaluated models across the four evaluation perspectives. 
% Table~\ref{tab:research_attention&hallucination} presents results for research attention and hallucination resistance, covering low- vs.\ high-attention genes for gene full name, chromosome location, and gene type, as well as hallucination resistance on synonyms and chromosome location. 
% Table~\ref{tab:knowledge_completeness&literature} summarizes performance on knowledge completeness and literature influence, including multi-answer tasks (related proteins, synonyms, and expression patterns) and Gene Ontology / functional summary questions with or without reference abstracts. 
% For single-choice and multiple-choice, scores are reported as mean~$\pm$~standard deviation over cyclic permutations of answer options. 
% In both tables, models marked with an asterisk (*) denote biomedical domain–specialized LLMs.

\section{Future Directions}
\label{appendix:future}

% While \bench{} provides comprehensive coverage of gene-level reasoning behaviors, it currently focuses on genes as isolated functional units.
% An important direction for future work is to extend the benchmark toward higher-order biological reasoning, such as regulatory relationships, gene-gene interactions, and pathway- or network-level contexts.
% Incorporating such structures may enable deeper assessment of how LLMs integrate structured biological knowledge with contextual evidence.
% Another promising direction is to explore the role of gene-centric reasoning in downstream biological interpretation workflows, including reference-based annotation and functional interpretation in large-scale cell atlas studies.
% Benchmarking how gene-level reliability propagates to higher-level biological analyses may further clarify the practical implications of LLM behavior in real-world biomedical settings.
% Finally, extending the benchmark to additional species, conditions, and emerging genomic resources could further enhance its coverage and applicability, supporting more robust evaluation of LLMs across diverse biological contexts.

While \bench{} covers gene-level reasoning behaviors, it currently treats genes as isolated functional units.
Future work can extend the benchmark to higher-order biological reasoning, including regulatory relationships, gene-gene interactions, and pathway- or network-level contexts.
Such extensions would better assess how LLMs integrate structured biological knowledge with contextual evidence.
Another direction is to examine how gene-level reliability affects downstream interpretation workflows, such as reference-based annotation and functional interpretation in large-scale cell atlas studies.
Finally, extending \bench{} to additional species, conditions, and emerging genomic resources would further improve its coverage and applicability.

\section{Online Resources and AI Tools Usage}
\label{appendix:online}

Figure~\ref{fig:online} presents the platform results and evaluation perspectives of \bench{}.
The platform supports filtering and ranking by model type and evaluation perspective.
We publicly release example questions, prompts, metric code, and a step-by-step notebook on GitHub\footnote{https://github.com/CNIC-DSL/SciHorizonGene}.
The evaluation dataset is available on Science Data Bank\footnote{https://www.scidb.cn/detail?dataSetId=4700d275bd5741958894d3739cbdc1dd\&version=V1}, and the official GO identifiers are available from the Gene Ontology website\footnote{\url{https://geneontology.org/docs/download-ontology/go-basic.obo}}.
Generative AI tools were used only as evaluation targets and for grammar checking.
All formulation, design, dataset construction, analysis, and conclusions were completed by the authors.

% Figure~\ref{fig:online} presents the evaluation results available on the platform, as well as the evaluation perspectives defined in \bench{}.
% The platform supports efficient filtering and ranking of results by model types and evaluation perspectives, enabling flexible and fine-grained analysis.
% We publicly release the evaluation resources of \bench{} to support reproducibility and further research.
% The evaluation results are updated on the platform, which supports filtering and ranking by model type and evaluation perspective.
% We provide example questions, prompts, metric computation code, and a step-by-step evaluation notebook on GitHub\footnote{https://github.com/CNIC-DSL/SciHorizonGene}.
% The evaluation dataset is publicly available on Science Data Bank\footnote{https://www.scidb.cn/detail?dataSetId=4700d275bd5741958894d3739cbdc1dd\&version=V1}, and the official GO identifiers used for metric computation are available from the Gene Ontology website\footnote{\url{https://geneontology.org/docs/download-ontology/go-basic.obo}}.
% Generative AI tools were used only as evaluation targets for benchmarking and for grammar checking.
% All problem formulation, experimental design, dataset construction, analysis, and conclusions were carried out by the authors.

\end{document}